\documentclass[epj,nopacs]{svjour}
\usepackage{graphicx} 
\usepackage{amsmath}
\usepackage{bm,amsfonts,amssymb}
\usepackage{rotating}
\usepackage{color}
\usepackage{amssymb}
\usepackage{cite}
\usepackage{enumerate}

\newcommand{\BE}{\begin{equation}}
\newcommand{\EE}{\end{equation}}

\newcommand{\One}{1\!\!1}

\begin{document}
\title{Second resonance of the Higgs field: \\ motivations, experimental
signals, unitarity constraints}
\author{Maurizio Consoli\inst{1} \and George Rupp\inst{2}}
\institute{
Istituto Nazionale di Fisica Nucleare, Sezione di Catania,
I-95123 Catania, Italy
\and
Centro de  F\'{\i}sica e Engenharia de Materiais Avan\c{c}ados,
Instituto Superior T\'{e}cnico, Universidade de Lisboa,
P-1049-001 Lisboa, Portugal
}
\date{Received: date / Revised version: date}
%
\abstract{Perturbative calculations predict that the effective
potential of the Standard Model should have a new minimum, well
beyond the Planck scale, which is much deeper than the electroweak
vacuum. So far, most authors have accepted the metastability
scenario in a cosmological perspective which is needed to explain
why the theory remains trapped in our electroweak vacuum but
requires to control the properties of matter in the extreme
conditions of the early universe. As an alternative, one can
consider the completely different idea of a non-perturbative
effective potential that, as at the beginning of the Standard Model,
is restricted to the pure $\Phi^4$ sector but is consistent with the
indications of the now existing analytical and numerical studies,
namely ``triviality'' and a description of SSB as weak first-order
phase transition. In this approach, the electroweak vacuum is now
the lowest energy state because, besides the state with mass
$m_h=125$~GeV, defined by the quadratic shape of the potential at
its minimum, there is a second much larger mass scale $(M_H)^{\rm
Theor} \sim 690(30)$ ~GeV associated with the zero-point energy
determining the potential depth. Despite its large mass, the heavier
state would couple to longitudinal $W$s with the same typical
strength as the low-mass state at 125~GeV and thus represent a
relatively narrow resonance mainly produced at LHC by gluon-gluon
fusion. Therefore, it is interesting that, in the LHC data, one can
find combined indications of a new resonance of mass $(M_H)^{\rm
comb} \sim 685 (10)$~GeV, with a statistical significance which is
far from negligible. Since this non-negligible evidence could become
an important new discovery with forthcoming data, we outline further
refinements of the theoretical predictions, that could be obtained
by implementing unitarity constraints, in the presence of fermion
and gauge fields, with the type of coupled-channel calculations
nowadays used in meson spectroscopy.}

\authorrunning{M.~Consoli and G.~Rupp}
\maketitle
\section{Introduction}
\label{intro}

The discovery at CERN \cite{discovery1,discovery2} of the
narrow scalar resonance with mass $m_h\!=\!$ 125 GeV and the consistency
of its phenomenology with the theoretical expectations for the Higgs
boson have confirmed spontaneous symmetry breaking (SSB) through
the Higgs field as a fundamental ingredient of current particle
physics. Nevertheless, in our opinion there may be room for improving on
the present description of symmetry breaking. The latter is based on a
classical double-well potential with perturbative quantum corrections,
say $V^{\rm (p)}(\phi)$, exhibiting a local minimum at
$\phi\!=\!v\sim 246$~GeV and with a quadratic shape fixed by
$m^2_h\!=\!(125\;\mbox{GeV})^2$.  The point is that, at large values of
$\phi$, this potential is well approximated as
$V^{\rm (p)}(\phi) \sim \lambda^{\rm (p)}(\phi)\,\phi^4$, in terms of the
perturbative scalar coupling $\lambda^{\rm (p)}(\phi)$ which
includes the effect of the gauge and fermion fields and becomes negative
beyond an instability scale $\phi_{\rm inst} \sim 10^{10}$~GeV. As a net
result, besides the local electroweak vacuum where
$V^{\rm(p)}(v)\sim -10^{8}\;\mbox{GeV}^4$, the true, absolute minimum of
this perturbative potential is at $v_{\rm true}\sim 10^{26}\div 10^{31}$~GeV
\cite{branchina,gabrielli} (depending on the approximations and the
exact values of the input parameters), with a much deeper potential
value $V^{\rm (p)}(v_{\rm true})\sim - (10^{100}\div 10^{120})$~(GeV)$^4$.

While it is reassuring that the most accurate calculation
\cite{degrassi} gives a tunneling time which is larger than the age
of the universe, still the idea of a metastable vacuum raises
several questions. For instance, the new minimum is much larger than
the Planck mass $M_P$ and the Planck scale is usually regarded as
the scale where gravity becomes strong. Thus, should the problem at
large $\phi \sim M_P$ be formulated in a curved space-time? In
this case, does the second minimum disappear? Here, due to various
uncertainties, there is no general consensus \cite{strumia,branchina2}
that gravitational physics at the Planck scale can become so strong to
stabilise the electroweak vacuum. On the contrary, the vanishing value
of the observed cosmological constant, on a particle-physics scale, could
imply that gravity remains weak \cite{gabrielli} at all energies without
introducing any threshold effect near $M_P$.

Thus, one has been considering the metastability scenario in a
cosmological perspective, because in an infinitely old universe even
an arbitrarily small tunneling probability would be incompatible
with our existence \cite{finlandesi}. However, given the extreme
conditions in the early universe, the survival of the tiny
electroweak minimum is somewhat surprising. As an example, before
the discovery of the resonance at 125~GeV, the authors of
Ref.~\cite{riotto} were concluding that, for
$114\;\mbox{GeV} \leq m_h \leq 130$~GeV and from the analysis of cosmological
perturbations, either we live in a very special and exponentially unlikely
corner or new physics must exist below $\phi_{\rm inst}\sim 10^{10}$~GeV.

As an alternative, one can consider the completely different idea of
a non-perturbative effective potential. Indeed, if SSB represents a
{\it non-perturbative} \/phenomenon, one may also try to describe it
non-perturbatively. Since a non-perturbative analysis can hardly
be carried out by retaining the full gauge and fermion structure of the
theory, as in the early days of the Standard Model, one could first
concentrate on the pure $\Phi^4$
sector. Nonetheless, in view of the substantial theoretical progress
over the past fifty years, this implies trying to describe SSB consistently
with the existing theoretical and numerical studies. Obtaining such a
description is a preliminary step in comparing the contributions of the
various sectors to vacuum stability and then, eventually, to conclude that
SSB is basically a phenomenon arising within pure $\Phi^4$ theory.

To this end, we will first focus on a one-component
$\Phi^4$ theory, i.e., without Goldstone bosons, and postpone
the case of several scalar fields to a later stage. The point is that the
one-component theory is a convenient laboratory, which already contains all
ingredients needed to describe SSB as a (weak) first-order phase transition.
This weak first-order picture has already been considered in
Refs.~\cite{Consoli:2020nwb,symmetry,memorial}, by first following the
indications of lattice simulations
\cite{lundow2009critical,Lundow:2010en,akiyama2019phase} and then
considering the known approximations to the effective potential that are
consistent with both this scenario and the basic ``triviality'' property of
the theory. Since these approximations, albeit physically equivalent, resum
to all orders different classes of diagrams, the resulting scheme can be
considered non-perturbative. In this way, one obtains a picture where the
(absolute) electroweak minimum of the potential can coexist with a quadratic
shape $V''_{\rm eff}(\phi\!=\!0) =m^2_\Phi>0$ which is very small yet still
positive. This leads to an intuitive picture of the broken-symmetry phase as
a condensate of physical quanta with mass $m_\Phi$ whose collective
self-interaction represents the {\it primary} \/sector that induces SSB.
For convenience of the reader, the essential points will be
summarised in this introduction.

The main new feature of this first-order description is that the
mass scale $M_H$ associated with the zero-point energy (ZPE), and which
determines the potential depth, is much larger than the mass scale $m_h$
defined by the quadratic shape of the potential at its minimum. This is
because, differently from the second-order scenario where the instability
of the symmetric phase is driven by a negative mass squared, ZPEs have now
to compensate for a tree-level potential that otherwise would have no
non-trivial minimum. The large difference of the two mass scales produces an
ambiguity in the definition of the vacuum field $v\sim 246$~GeV that has no
counterpart in the usual perturbative approach. Resolving this ambiguity
requires a Renormalisation-Group (RG) analysis of the SSB phenomenon, which
is also essential to conclude that the contribution of the gauge and fermion
fields to vacuum stability can be considered a small radiative correction.

Such an RG analysis is needed because the quartic coupling
$\lambda\!=\!\lambda(\mu)$ associated with the self-interaction of the
primary scalar sector is positive definite and exhibits a Landau pole
$\Lambda$. This is linked to the distance $r_0\!\sim\!\Lambda^{-1}$ at which
the elementary quanta feel a hard-core repulsion, while at a finite scale
$\mu$ one has $\lambda(\mu)\!\sim\!L^{-1}$ in terms of
$L\!=\!\ln (\Lambda/\mu)$. Now, since for any non-zero $\lambda$ there is a
Landau pole $\Lambda$, one can improve the description by considering the
set of theories ($\Lambda$,$\lambda$), ($\Lambda'$,$\lambda'$),
($\Lambda''$,$\lambda''$), \ldots with larger and larger Landau poles,
smaller and smaller low-energy couplings at $\mu$, but all having the same
depth of the potential at the minimum $\phi\!=\!\phi_v$, i.e., with the same
vacuum energy $\mathcal{E}\!=\!V_{\rm eff}(\phi_v)$ as determined by the
equation
\BE
\label{CSground}
\left(\Lambda\frac{\partial}{\partial\Lambda} +
\Lambda\frac{\partial \lambda}{\partial\Lambda}\frac{\partial
}{\partial \lambda}\right){\cal E}(\lambda,\Lambda)\; =\; 0 \; .
\EE
This requirement derives from imposing RG invariance of the effective
potential in the three-dimensional space ($\phi$, $\lambda$, $\Lambda$)
\cite{Consoli:2020nwb,symmetry,memorial}, namely
\BE
\label{CSveff} \left(\Lambda\frac{\partial}{\partial\Lambda} +
\Lambda\frac{\partial \lambda}{\partial\Lambda}\frac{\partial
}{\partial \lambda}  + \Lambda\frac{\partial
\phi}{\partial\Lambda}\frac{\partial }{\partial \phi}
\right)  V_{\rm eff}(\phi,\lambda,\Lambda) \; = \; 0 \; ,
\EE
and can in principle also allow one to handle the $\Lambda\!\to\!\infty$
limit.\footnote{The primary scalar sector is assumed to induce
SSB and determine the vacuum structure. In a quantum field theory, invariance
under RG transformations is the usual method to remove the ultraviolet cutoff
or, alternatively, to minimise its influence on observable quantities.}
Now, in those known approximations to the effective potential
that are consistent with ``triviality'' and a weak first-order scenario
of SSB, in terms of the ZPE mass scale $M_H$ one finds
${\cal E}=V_{\rm eff}(\phi_v) \sim -M^4_H$. By Eq.~(\ref{CSground}) this
means that $M_H\sim \Lambda \exp (-1/\lambda)$ is an RG-invariant mass scale
$I_1=M_H$ or, equivalently, that $M_H$ is the scale within the logarithm
$L=\ln (\Lambda/M_H)$ in the low-energy coupling $\lambda \sim  L^{-1}$.
However, for the same reason the relation
$M^2_H \sim \lambda \phi^2_v \sim \phi^2_v L^{-1} $ implies that
$\phi_v\sim M_H L^{1/2}$ {\it cannot} \/represent the Fermi scale, which is
always assumed to be a cutoff-independent quantity. Analogously, the
quadratic shape of the potential, which is obtained by twice differentiating
the potential with respect to the cutoff-dependent $\phi_v $, will be
much smaller than $M_H$, namely
$m^2_h=V''_{\rm eff}(\phi_v)\sim M^2_H L^{-1}$.

A solution to this problem can be found by noticing that in the RG
analysis there is a second invariant $I_2$
\cite{Consoli:2020nwb,symmetry,memorial}, related to a particular
normalisation of the vacuum field and in terms of which the minimisation
of the energy can be expressed as $I_1= K I_2$, with $K$ being a
cutoff-independent constant. This $I_2$ is then the candidate to represent
the weak scale, i.e., $I_2\!=\!v\sim 246$~GeV, giving $M_H=K v$. Since a
cutoff-independent $v$ should scale as
$v\sim\phi_v\sqrt{\lambda}\sim\phi_v L^{-1/2}$ and $m_h\sim M_H L^{-1/2}$,
the natural relation between $v$ and $\phi_v$ thus becomes
\BE
\label{fundv}
v \; = \; \frac{m_h}{M_H} \, \phi_v  \; .
\EE
This way, with
\BE
\label{same0}
\frac{3 M^2_H}{\phi^2_v} \; = \; \frac{3 m^2_h}{v^2} \; ,
\EE
the two fourth-order scalar couplings have the same value
at the Fermi scale $\mu\!\sim\!v$, namely
\BE
\label{same}
\lambda(v) \; = \; \lambda^{\rm (p)}(v) \; = \; 3 m^2_h/v^2 \; ,
\EE
while still behaving quite differently at very large $\mu$.

Now, if the higher-momentum mass scale $M_H$, associated with ZPEs, differs
non-trivially from the zero-momen\-tum mass scale $m_h$ defined by the
quadratic shape of the potential at its minimum, the Higgs field propagator
should deviate from a standard one-pole structure. The existence of these
deviations has been checked with lattice simulations of the propagator, which
have also confirmed the expected scaling trend $M^2_H \sim m^2_h L$
\cite{Consoli:2020nwb}. Thus one arrives at the conclusion that, besides the
known resonance with $m_h\!=\!125$~GeV, there should be a second resonance with
a much larger mass, which by combining numerical and analytical relations
can be estimated to have a value $M_H\!\sim\!700$~GeV. With such a
large $M_H$, the ZPEs of all known gauge and fermion fields would represent
a small radiative
correction,\footnote{To this end, it is crucial that the gauge and
fermion fields get their masses from the corresponding couplings times
$v\sim 246$~GeV (and {\it not} \/times the cutoff-dependent $\phi_v$).}
so that, by restricting to experiments in a region around the Fermi scale,
say a few TeV, the different evolution of $\lambda^{\rm (p)}(\mu)$ and
$\lambda(\mu)$ at asymptotically large $\mu$ should remain unobservable.
The crucial check of our picture is then the necessity to experimentally
observe the second resonance. Its discovery would mean that SSB is a
phenomenon originating within the primary scalar sector, namely from the
collective self-interaction of the basic quanta of the symmetric phase.

After outlining in Sec.~2 the weak first-order picture of SSB, we will
consider in Sec.~3 the basic phenomenology of the second resonance.
In Sec.~4, we will then enlarge the data sample considered in
Refs.~\cite{signals,ichep,CFF,universe} by including other LHC data that
strengthen the evidence of a new resonance in the expected mass range. As we
will show, the present non-negligible statistical evidence could become an
important discovery by adding new data. In view of these possible future
developments, we will illustrate in Sec.~5 the basic ingredients of a
coupled-channel calculation, which could be useful to further refine
the theoretical predictions for the mass and width of the
hypothetical new resonance when interacting with the gauge and
fermion sectors of the Standard Model. Section~6 will be devoted to
our conclusions, besides some remarks about the present agreement
between the Higgs-mass parameter extracted indirectly from radiative
corrections and the value $m_h=125$~GeV directly measured at the
LHC.

\section{SSB in a $\Phi^4$ theory}

\subsection{Preliminaries}

Let us start from scratch with the type of scalar potential reported in
the review of the Particle Data Group (PDG) \cite{PDG2022}:
\begin{equation}
\label{VPDG}
V_{\rm PDG}(\phi) \; = \; -\frac{ 1}{2} m^2_{\rm PDG}\,
\phi^2 + \frac{ 1}{4!}\lambda_{\rm PDG}\,\phi^4 \; .
\end{equation}
By fixing $m_{\rm PDG}\sim 88.8$~GeV and $\lambda_{\rm PDG}\sim0.78$,
this has a minimum at $|\phi|=v \sim 246$~GeV and a second derivative
$V''_{\rm PDG}(v)\equiv m^2_h=$~(125 GeV)$^2$ (one is adopting here the
identification $m^2_h= V''_{\rm PDG}(v)=|G^{-1}(p=0)|$ in terms of
the inverse, zero-momentum propagator).

In Eq.~(\ref{VPDG}), one is assuming a double-well potential with
suitably chosen mass and coupling. The instability of the symmetric
vacuum at $\phi=0$ is then traced back to the condition
$V''_{\rm PDG}(\phi=0)=-m^2_{\rm PDG}<0$, which characterises SSB as a
second-order phase transition. This traditional idea of a ``tachyonic''
mass term at $\phi=0$, however, is not the only possible
explanation. As in the original analysis by Coleman and Weinberg
\cite{Coleman:1973jx}, SSB could originate from the ZPE in the classically
scale invariant limit $V''_{\rm eff}(\phi=0) \to 0^+$. In this case, if the
quanta of the symmetric phase have a tiny physical mass
$m^2_\Phi \equiv V''_{\rm eff}(\phi=0)>0$ below some critical value $m^2_c$,
the symmetric phase could be ``locally'' stable but become ``globally''
unstable. By lowering the mass below $m^2_c$, the absolute minimum
of the effective potential would then discontinuously jump from
$\phi=0$ to $\phi\neq 0$ and SSB would represent a first-order phase
transition.

In order to understand how subtle the issue can be and get some
intuitive insight, let us consider the following toy model:
\begin{equation}
\label{toy}
V_{\rm toy}(\phi) \; = \; \frac{ 1}{2} m^2 \phi^2 + \frac{
1}{4!}\lambda \phi^4 \left(1 + \epsilon\ln\frac{\phi^2}{\mu^2}\right) \;,
\end{equation}
where $\mu$ is some mass scale and $\epsilon$ is a small,
positive parameter (the quantum-theory case is with $\epsilon \sim
\lambda$, but in this toy model we treat $\epsilon$ as a separate
parameter). For $\epsilon=0$, where  $V_{\rm toy}(\phi)$ reduces to
the classical potential $V_{\rm cl}(\phi)$, by varying the $m^2$
parameter there is a second-order phase transition at $m^2=0$ .
However, for any $\epsilon>0$, no matter how small, one has a
first-order transition, occurring at a positive $m^2$. The size of the
critical $m^2_c$ is exponentially small, viz.\ $\mu^2\exp(-1/\epsilon)$,
meaning that an infinitesimally weak first-order transition can become
indistinguishable from a second-order transition, unless one looks on a
fine enough scale.

We emphasise that this idea of SSB as a weak first-order phase
transition in $\Phi^4$ theories finds support in lattice simulations
\cite{lundow2009critical,Lundow:2010en,akiyama2019phase}. To that
end, one can just look at Fig.~7 in Ref.~\cite{akiyama2019phase},
where the data for the average field at the critical temperature
show the characteristic first-order jump and not the smooth
second-order trend. This agreement with lattice simulations is a
good motivation to further explore the physical implications of a
first-order scenario.\footnote{We note that conflicting indications
have more recently been reported in Ref.~\cite{lundow22}. These
authors object to the traditional view that the Ising model and
$\Phi^4$ theory (at finite bare coupling) belong to the same
universality class. Thus, a second-order phase transition, as in
standard RG-improved perturbation theory, would
not be ruled out. Nonetheless, the Ising limit, with a lattice
coupling at the Landau pole, is known to saturate the triviality
bound in $\Phi^4$ theory. Namely, at any fixed non-zero value of the
renormalised coupling, it represents the best approximation to the
continuum limit \cite{gliozzi}, a remark that is certainly relevant
for lattice simulations of a quantum field theory. Furthermore, a
weak first-order transition can become asymptotically second-order
in the continuum limit and our toy model in Eq.~(\ref{toy})
illustrates how delicate the issue can become numerically in the
$\epsilon \to 0$ limit. Finally, as we shall see in Subsec.~2.2
the weak first-order scenario of SSB in $\Phi^4$ gathers {\it
additional} \/motivations when considering the class of
approximations to the effective potential that are consistent with
the basic ``triviality'' of the theory.}

From a purely physical point of view, the underlying rationale for a
tachyonic mass term at $\phi=0$ reflects the basic prejudice that
$\Phi^4$ is an interaction that is always repulsive. In fact, with
a purely repulsive interaction, any state made of massive physical
particles, would necessarily have an energy density that is higher
than the trivial empty vacuum at $\phi=0$. However, as discussed in
Ref.~\cite{Consoli:1999ni}, the $\Phi^4$ interaction is {\it not}
\/always repulsive. The interparticle potential between the basic
quanta of the symmetric phase, besides the $+\lambda\,
\delta^{3}(\bf r)$ tree-level repulsion, contains a
$-\lambda^2\,e^{-2 m_\Phi r}/r^3$ attraction, which originates in the
{\it ultraviolet-finite} \/part of the one-loop diagrams and whose range
becomes longer and longer in the $m_\Phi \to 0$
limit.\footnote{Starting from the scattering-matrix element $\cal M$
obtained from Feynman diagrams, one can construct an interparticle
potential that is basically the three-dimensional Fourier transform
of $\cal M$; see Refs.~\cite{Feinberg:1968zz,Feinberg:1989ps}.}
Due to the qualitative difference between the two effects and in order to
consistently include higher-order effects, one should rearrange the
perturbative expansion by symmetrically renormalising {\it both} \/the
contact repulsion and the long-range attraction as discussed by Stevenson
\cite{Stevenson:2008mw}. In this way, by taking into account both
effects, a calculation of the energy density indicates that, for
positive and small enough $m_\Phi$, the attractive tail dominates.
Then, the lowest-energy state is not the trivial, empty vacuum with
$\phi=0$, but a state with $\phi\neq 0$ and a Bose condensate of
symmetric-phase quanta in the $\vec{k} = \vec{0}$
mode.\footnote{This first-order scenario is implicit in 't Hooft's
description of SSB \cite{thooft}: \em ``What we experience as empty space
is nothing but the configuration of the Higgs field that has the lowest
possible energy. If we move from field jargon to particle jargon, this
means that empty space is actually filled with Higgs particles. They have
Bose condensed''. \em This clearly refers to real, physical quanta.
Otherwise, in a second-order picture, what Bose condensation
would there be at all?}

\subsection{``Triviality'' and the effective potential}
\begin{figure}[htb]
\includegraphics[trim = 0mm 0mm 0mm 0mm,clip,width=8.6cm]
{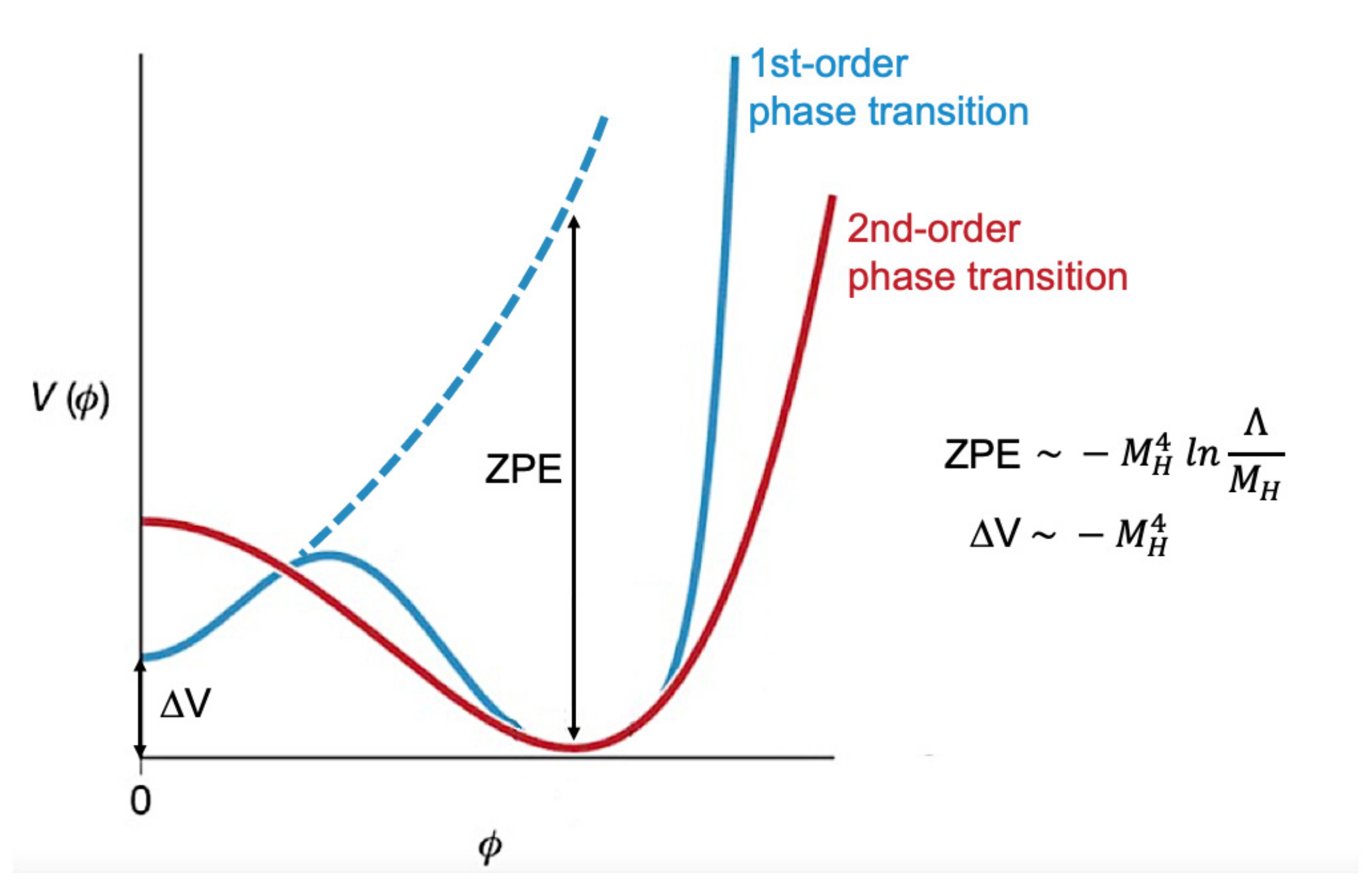}
\caption{An
intuitive picture that illustrates the crucial role of the ZPE in a
first-order scenario of SSB. Differently from the standard
second-order picture, it has to compensate for a tree-level
potential with no non-trivial minimum.}
\label{phase}
\end{figure}
In spite of these interesting aspects, one could still wonder about
different observable consequences. After all, the phenomenology of
the broken-symmetry phase should only depend on the potential near
the true minimum (so not at $\phi=0$) and, in principle, nothing
prevents it locally from having exactly the same shape as in
Eq.~(\ref{VPDG}). To get some insight, let us look at Fig.~\ref{phase}.
This intuitively illustrates that, if $V''_{\rm eff}(\phi=0)> 0$, ZPEs
are expected to be much larger than in a second-order picture. In
the latter case, in fact, SSB is driven by the negative mass squared
at $\phi=0$, while now ZPEs have to overwhelm a tree-level potential
that otherwise would have no non-trivial minimum. But what do we
exactly mean by saying that ZPEs have to be {\it  much larger}? The
answer is that, now, the ZPE mass scale $M_H$ is much larger than
the mass scale $m_h$ defined by the quadratic shape of the effective
potential at the minimum. To fully understand this crucial issue let
us first recall that this large size of ZPEs induced Coleman and
Weinberg to expect that the weak, first-order scenario could only
work in the presence of gauge bosons. In a pure $\Phi^4$ theory, SSB
would require to compensate the positive $\lambda \phi^4$ tree-level
potential with a negative $\lambda^2\phi^4 \ln(\phi^2/\Lambda^2)$
one-loop contribution, a requirement that lies outside a standard
loop-expansion perspective. Instead, they concluded that with gauge
bosons the corresponding one-loop contribution $g^4_{\rm
gauge}\phi^4 \ln(\phi^2/\Lambda^2)$ could well represent the needed
driving mechanism if $\lambda \sim g^4_{\rm gauge}$.

Nevertheless, there is a way to rearrange things in the effective
potential and consistently describe SSB as first-order transition. The
standard perspective, which is behind the idea of the perturbative
potential  $V^{\rm (p)}(\phi) \sim \lambda^{\rm (p)}(\phi)\phi^4$,
considers the one-loop contribution as simply renormalising the
coupling $\lambda$ in the classical potential
\begin{equation}
\frac{\lambda}{4!}\,\phi^4 \; \to \; \frac{\lambda}{4!}\,\phi^4
\left\{\!1-\frac{3\lambda}{32\pi^2} \ln \left(\frac{2\Lambda^2
}{\lambda \phi^2}\right) \right\} \; .
\end{equation}
Therefore, including the higher-order leading logarithmic
terms, i.e., by replacing
\begin{equation}
1-x \to 1-x +x^2 -x^3 \pm \ldots \; = \; 1/(1+x) \; ,
\end{equation}
the one-loop minimum would disappear.

But, as emphasised by Stevenson \cite{Stevenson:2008mw}, the
qualitatively different nature of the two basic terms in the
inter-particle potential between the quanta of the symmetric phase
has a definite counterpart in the structure of the effective
potential. Here, the positive $\lambda\phi^4$ background originates
from the +$\lambda \delta^{3}(\bf r)$ short-range repulsion and the
negative ZPE from the long-range $-\frac {\lambda^2}{r^3}$
attraction. This observation suggests to consider the equivalent
reading of the one-loop potential as the sum of a classical
background plus ZPEs of free-field fluctuations with mass squared
$M^2(\phi)= \lambda \phi^2/2$
\begin{equation}
V_{\mbox{\scriptsize 1-loop}}(\phi)  =
\frac{\lambda\phi^4}{4!} -\frac{M^4(\phi) }{64 \pi^2} \ln
\frac{\Lambda^2 \sqrt{e} } {M^2(\phi)} .
\end{equation}
Since this type of structure is
also recovered in higher-order approximations, the simple one-loop
potential can also admit a non-perturbative interpretation as the
{\it prototype} \/of a class of calculations with the same basic
structure up to a redefinition of {\it both} \/the classical background
and the mass parameter $M(\phi)$.

This is explicitly illustrated by the Gaussian effective potential
\cite{gaussian} which re-sums all one-loop bubbles  and preserves
the same structure up to terms that vanish when $\Lambda \to
\infty$:
\begin{equation}
\lambda \; \to \; \lambda_G(\phi) \; = \; \frac{\lambda } {1 +
\frac{\lambda}{16 \pi^2} \ln \frac {\Lambda}{ M_G(\phi)} } \; ,
\end{equation}
\begin{equation}
M^2(\phi) \; \to \; M^2_G(\phi) \; = \; \frac{\lambda_G(\phi)\phi ^2}{2} \; ,
\end{equation}
\begin{equation} \label{vgauss}
V_{\mbox{\scriptsize 1-loop}}(\phi)  \to  V_G(\phi) =
\frac{\lambda_G(\phi)\phi^4}{4!} -\frac{M^4_G(\phi) }{64 \pi^2} \ln
\frac{ \Lambda^2 \sqrt{e} } {M^2_G(\phi)}  .
\end{equation}
The agreement between the one-loop and Gaussian effective potential
has to be emphasized, because it gives further insight into the
``triviality'' of  $\Phi^4$. If, in the continuum limit, all
interaction effects have to be effectively reabsorbed into the first
two moments of a Gaussian distribution, meaningful approximations to
the effective potential should be physically equivalent to the
one-loop result, i.e., given again by some classical background +
ZPE of free-field fluctuations with some $\phi-$dependent mass
\footnote{Further examples of such ``triviality-compatible''
approximations are the post-gaussian calculations
\cite{Stancu:1989sk,Cea:1996pe}. These have a propagator determined
variationally and can become arbitrarily complex.}.

For this reason, as anticipated in the Introduction,
the two approximations considered above produce
equivalent results. Namely, by using the same notation $\phi=\phi_v$
for the minimum of the effective potential, either when
$V_{\rm eff}(\phi)=V_{\mbox{\scriptsize 1-loop}}(\phi)$ or when
$V_{\rm eff}(\phi)=V_G(\phi)$, and by denoting $M_H$ as the value of
$M(\phi_v)$ or of $M_G(\phi_v)$, one finds
$V_{\rm eff}(\phi_v)=-M^4_H/(128\pi^2)$ and
$M^2_H \sim \lambda\phi^2_v \sim \Lambda^2 \exp (-1/ \lambda)$.
The important point is that the vacuum energy
is an RG-invariant quantity satisfying Eq.~(\ref{CSground}),
so that $M_H$ is $\Lambda$-independent and one can express $\lambda \sim
{L}^{-1}$, with $L=\ln (\Lambda/M_H)$.

However, for the same reason $\phi_v\sim M_H {L}^{1/2}$ cannot
represent the Fermi scale, which is always assumed to be a cutoff-independent
quantity. Nevertheless, since
$m^2_h=V''_{\rm eff}(\phi_v) \sim \lambda^2 \phi^2_v\sim  M^2_H  {L}^{-1}$
as anticipated in the Introduction, through Eq.~(\ref{fundv}) one can
introduce a vacuum field $v$ that scales uniformly with $M_H$ and is therefore
cutoff-independent. This way, one finally obtains the following pattern of
scales \cite{Consoli:2020nwb,symmetry,memorial}:
\begin{equation}
\label{scale}
\lambda  \sim  {L}^{-1}  , \;
m^2_h  \sim  v^2 \, {L}^{-1}  , \;
M^2_H  \sim  L \, m^2_h  =  K^2 v^2  ,
\end{equation}
where $K$ is a cutoff-independent constant and
\begin{equation}
\phi^2_v \; \equiv \; Z_\phi v^2 \; , \;\;\;{\rm with} \;\;\;
Z_\phi \; = \; (M_H/m_h)^2 \sim L \; .
\end{equation}
Then, $M_H$ and $v$ will emerge as the two invariants
\cite{Consoli:2020nwb,symmetry,memorial} $I_1=M_H$
and $I_2=v$ associated with the analysis of the effective potential
in the $(\lambda,\phi,\Lambda$) three-dimensional space, in terms of
which the absolute-minimum condition can be expressed as $I_1= K I_2$.
For this reason, $v$ represents the natural candidate to
represent the weak scale $v\sim 246$ GeV. Note that in perturbation
theory in the standard second-order scenario, where $m_h\sim M_H$,
there is no $v$-$\phi_v$ distinction. Instead, here it is a
consequence of minimising the effective potential and the strong
cancelations between formally higher-order and tree-level terms.
Further implications of this two-mass structure will be illustrated
in the following two subsections.

\subsection{The coexistence of $m_h$ and $M_H$}

To further sharpen the meaning of $m_h$ and $M_H$, let us recall
that the ZPE is (one half of) the trace of the logarithm of the
inverse propagator $G^{-1}(p)=(p^2-\Pi(p))$. Therefore, in a
free-field theory, where $V_{\rm free}(\phi )= m^2 \phi^2/2$ and
$|\Pi(p)| = |\Pi(p=0)|= m^2 $ one finds
\begin{equation}
({\rm ZPE})^{\rm free}  =  \frac{1}{2} \int\!\!\frac{d^4
p}{(2\pi)^4} \ln (p^2 + |\Pi(p=0)|).
\end{equation}
Instead, in the presence of interactions when in general $\Pi(p)
\neq \Pi(p=0)$, things are not so simple. On the one hand, the
derivatives of the effective potential produce (minus) the $n$-point
functions at zero external momentum, so that, by defining $\phi_v$
as the minimum of $V_{\rm eff}(\phi )$, one gets
\begin{equation}
m^2_h \; \equiv \; V''_{\rm eff}(\phi_v) \; = \;
|\Pi(p=0)| \; = \; |G^{-1}(p=0)| \; .
\end{equation}
On the other hand, ZPEs contribute to the effective potential but
are {\it not} \/a pure zero-momentum quantity. Therefore, after
subtracting constant terms and quadratic divergences, one can write
at the minimum
\begin{eqnarray}
\label{general}
{\rm ZPE} & \; \sim \; & -\frac{1}{4}
\int^{p_{\rm max}}_{p_{\rm min}}\!\frac{d^4 p}{(2\pi)^4}
    \frac{\Pi^2(p)}{p^4}  \; \sim \; \nonumber \\[1mm]
&&-\frac{ \langle \Pi^2(p)\rangle }{64\pi^2}
\ln\frac{p^2_{\rm max}}{p^2_{\rm min}} \; \sim \;
-\frac{M^4_H}{64\pi^2} \ln\frac{\Lambda^2 }{M^2_H} \; .
\end{eqnarray}
This shows that $M^2_H$, effectively
including the contribution of the higher momenta, reflects a typical
average value $|\langle \Pi(p)\rangle| $ at non-zero $p$. In
perturbation theory, where $ \Pi(p) \sim \Pi(p=0) $ up to small
corrections, one finds $M_H \sim m_h$.  On the other hand, if
$M_H \gg m_h$, {\it there must be} \/a non-trivial difference between
$p = 0$ and $p\neq 0$, with deviations from a standard one-mass
propagator.

Discussing the $p$ dependence requires to switch from
the effective potential to the effective action within the same
class of approximations as used for the effective potential. For our
scope, the relevant approximation is the Gaussian Effective Action
(GEA), worked out by A.~Okopinska \cite{okopinska} in the
$O(N)$-symmetric case. After setting $N=1$ and replacing $\lambda
\to \lambda/4!$ in the $\Phi^4$ interaction term, from Eqs.~(15) and
(16) in Ref.~\cite{okopinska} one finds the optimal Gaussian mass at
the SSB minima $\phi=\pm \phi_v$ \BE \Omega^2(\phi_v) \; = \;
\lambda \phi^2_v /3 \; \equiv \; M^2_H \EE and an inverse scalar
propagator \BE \label{propagatorGEA} G^{-1}(p) \; = \; p^2 +  M^2_H
A(p,M_H) \; , \EE with \BE \label{Ap} A(p,M_H) \; = \; \frac{
1-J(p,M_H) }{\displaystyle 1+\frac{J(p,M_H)}{2}} \EE and \BE
\label{JpM} J(p,M) \; = \; \lambda\int\frac{d^4k }{(2 \pi)^4} \,
\frac {1}{(k^2 + M^2)[(k+p)^2 + M^2]} \; . \EE To understand the
above results in terms of a diagrammatic expansion, with scalar
potential $U(\phi)=m^2_B \phi^2/2 + \lambda\phi^4/4!$, let us adopt,
at any $\phi$, the following two-step procedure :
\begin{enumerate}[1)]
\item
first re-absorbing all momentum-independent one-loop tadpoles into a
mass $\Omega(\phi)$;
\item
then re-summing all (non-tadpole) one-loop bubbles with mass $\Omega(\phi)$.
\end{enumerate}
Step 1) corresponds to the self-consistent equation for the Gaussian
variational mass parameter
\BE
\Omega^2(\phi) \; = \; m^2_B + \frac{ \lambda \phi^2 }
{2} + \frac{ \lambda } {2} I_0[\Omega(\phi)] \; ,
\EE
where
\BE
\label{I_0}
I_0(\Omega) \; = \; \int\frac{d^4k}{(2\pi)^4} \, \frac{1}{k^2+\Omega^2} \; .
\EE
After that, step 2) amounts to considering the propagator series
\begin{eqnarray}
\lefteqn{G^{-1}(p) \; = \; } \nonumber \\
&&\hspace*{-3mm}p^2+\Omega^2-\frac{\lambda\phi^2}{2}J(p,\Omega) \left\{
1\!-\!\frac{J(p,\Omega)}{2}\!+\!\frac{J^2(p,\Omega)}{4}\!+\ldots\right\}=
\nonumber \\
&&\hspace*{-3mm}p^2+\Omega^2-\frac{\lambda\phi^2}{2}
\frac{J(p,\Omega)}{1+ J(p,\Omega)/2}\;,
\label{propseries}
\end{eqnarray}
where $\Omega= \Omega(\phi)$ everywhere. Since at the minimum
$\phi=\phi_v$ of the Gaussian effective potential
$\Omega^2(\phi_v)=\lambda\phi^2_v/3=M^2_H$, one then obtains
Eqs.~(\ref{propagatorGEA}) and (\ref{Ap}), where $J(p,M_H)$, in
terms of the parameter $\epsilon=\lambda/16\pi^2$, admits the two
equivalent expressions \BE J(p,M_H) = \epsilon
\left\{\ln\frac{\Lambda^2}{p^2} - \int^1_0 \! dx \, \ln\left[x(1-x)
+\frac{M^2_H} { p^2} \right]\right\} \EE and \BE J(p,M_H) = \epsilon
\left\{\ln\frac{\Lambda^2}{M^2_H} - \int^1_0 \! dx \,
\ln\left[1+x(1-x) \frac{p^2} { M^2_H} \right]\right\} . \EE To study
the propagator, we will assume the previous perspective of a cutoff
theory with a coupling  $\lambda=\lambda(\mu)$ vanishing as $[\ln
(\Lambda/\mu)]^{-1}$ in the continuum limit $\Lambda \to \infty$ and
the same Coleman-Weinberg scenario of a weak first-order phase
transition with mass parameter $\Omega(\phi=0)=\Omega_0 \to 0^+$, so
that SSB originates in dimensional transmutation from a nearly
scale-invariant theory. As anticipated, the Gaussian energy density
at the minimum maintains exactly the same form \BE V_{\rm
eff}(\phi_v )=V_G(\phi_v)= - \frac{ M^4_H} {128 \pi^2} \EE as for
the Coleman-Weinberg one-loop potential, with an $M_H$-$\Lambda$
relation in terms of $L=\ln (\Lambda/M_H)$ that now reads \BE
J(0,M_H)= 2 \epsilon L = 1-\epsilon \; . \EE By introducing the
dimensionless parameter $z=p^2/\Lambda^2$, the deviation from the
Gaussian mass can then conveniently be expressed as \BE G^{-1}(p) \;
= \; p^2+  M^2 (z,\epsilon) \; , \EE where \BE M^2(z,\epsilon) \; =
\; M^2_H \, A(z,\epsilon) \EE and \BE A(z,\epsilon) \; = \;
\frac{1-J(z,\epsilon)}{\displaystyle1+\frac{J(z,\epsilon)}{2}} \; .
\EE Here, the function $J(z,\epsilon) $ is given by \BE
J(z,\epsilon) \; = \; \epsilon\left[\ln \frac{1} { z} -
h(a^2)\right] \; , \EE with \BE a^2 \; = \; \frac{M^2_H} { p^2} \; =
\; \frac{1}{z} \exp\left[- \frac{(1-\epsilon)}{\epsilon} \right] \;
, \EE \BE h(a^2) \; = \; -2 + \ln(a^2) + \Delta
\ln\frac{\Delta+1}{\Delta-1} \; , \EE and \BE \Delta \; = \; \sqrt
{1 + 4a^2} \; . \EE Thus, for $\epsilon \ll 1$,  we get a
$A(0,\epsilon)\sim (2/3)\epsilon$ and a zero-momentum mass \BE
\label{mh1} m^2_h \; \equiv \; M^2_H A(0,\epsilon)
 \sim  \; M^2_H \, L^{ -1} \ll M^2_H \, \; , \EE while for
$\epsilon \ln (1/z) \ll 1$ we find $M(z,\epsilon)\sim M_H$. This
larger-momentum limit is better illustrated by introducing a small
parameter $\delta \ll 1$ and exploring the condition \BE
\label{zdelta}|1 -A( z,\epsilon)| \; < \; \delta \; . \EE For
instance, by requiring a mass $M(z,\epsilon)$ that differs from
$M_H$ by less than 5\% for any $z>\bar z$, we get $\delta \sim 0.1$
and the following pairs of values: i) $\bar z \sim 0.01$ for
$\epsilon=0.01$, ii) $\bar z \sim 10^{-5}$ for $\epsilon= 5\times
10^{-3} $, iii) $\bar z \sim 10^{-29}$ for $\epsilon= 1\times
10^{-3} $, \ldots\ . In general, $\bar z \sim \exp
(-2\delta/3\epsilon) $. This confirms that, for $\epsilon \ll 1$,
the function $|\Pi(p)|= M^2(z,\epsilon)$ in Eq.~(\ref{general}) is
close to $M^2_H$ nearly everywhere in the range $0 < z < 1$.
Finally, in the continuum limit where $\epsilon \to 0$, one finds
$M(z,\epsilon)=M_H$ at all but non-zero $p$, besides a discontinuity
at $p=0$. 

This peculiarity of the $p=0$ state could have been deduced on a
purely hypothetical basis without any specific calculation, by just
using the mentioned ``triviality'' property of $\Phi^4$ theory in
four dimensions (4D). Indeed, requiring a Gaussian structure of
Green's functions in the continuum limit of the cutoff theory does
not forbid a first moment $\langle \Phi \rangle \neq 0$. However, it
requires a continuum limit with a free-field {\it connected}
\/propagator $G(x-y)$, namely with Fourier transform
$G^{-1}(p) = (p^2 + M^2_H)$ for {\it any} \/$p_\mu \neq 0$. While this
leaves open the meaning of $G^{-1}(p=0)$, we observe that the zero-measure
set $p_\mu=0$ is transformed into itself under 4D Euclidean
rotations (or under the Lorentz Group in Minkowski space).
Therefore, a discontinuity at $p_\mu=0$ is a logical possibility to
reconcile SSB and ``triviality'' in the continuum limit of $\Phi^4$.
This apparently negligible discontinuity is the crucial difference
with respect to a standard continuum limit, seen as a totally
uninteresting massive free-field
theory.\footnote{There is a nice
analogy with non-relativistic quantum mechanics \cite{huang} when
solving the Schr\"odinger equation with a repulsive $\delta({\bf r})$
potential in three dimensions. By considering the $\delta$
potential as the limit of a sequence of well-behaved potentials of
smaller and smaller range, the condition that the wave function
vanish at ${\bf r}=0$ is automatically satisfied by all partial
waves except $S$-waves. For $S$-waves it cannot be satisfied if one
requires continuity at the origin. In this case, there would be no
$S$-waves and the solutions of the equation would not form a
complete set. But a discontinuity at ${\bf r}=0$ is acceptable,
because the potential is singular there. Therefore, despite the
vanishing of all phase shifts, $S$-wave states are not entirely free
due to the discontinuity at ${\bf r}=0$.}

\subsection{Lattice simulation of the propagator and the value of $M_H$}

The one-loop and Gaussian approximations to the effective potential
we have considered, both consistent with the basic ``triviality'' of
$\Phi^4$ and a weak first-order picture of SSB, indicate the
existence of deviations from a standard single-particle propagator.
Indeed, the Euclidean zero-momentum value $G^{-1}(p=0) \equiv m^2_h$
vanishes proportionally to $L^{-1}$ in units of the higher-momentum
scale $M^2_H$ associated with the ZPE. However, the numerical
coefficient $c_2$ that describes the logarithmic slope, say $M^2_H
\sim m^2_h L (c_2)^{-1}$, is different in the two cases. A similar result
was found in the post-Gaussian calculation of Ref.~\cite{Stancu:1989sk}.
The final expressions take exactly the same mathematical form as in the
Gaussian approximation, with only some numerical 
changes in the coefficients of the divergent logs. Since
lattice simulations are considered a reliable non-perturbative
approach, numerical simulations were thus performed
\cite{Consoli:2020nwb} in order to find the best approximations to a
free-field propagator and compute $m_h$ from the $p\to 0$ limit of
$G(p)$ and $M_H$ from its behaviour at higher $p^2$. In
this way one could first check the expected logarithmic trend and
then extract $c_2$. Simulations were performed in the Ising limit of
the theory, i.e. with a lattice coupling at the Landau pole. 
For any non-zero value of the renormalized coupling, this is known to
saturate the triviality bound and to provide the best approximation to
the continuum limit \cite{gliozzi}. The broken-symmetry phase then
corresponds to values of the hopping parameter $\kappa$ that are larger
than a critical value $\kappa_c=0.074848(2)$ \cite{Stevenson2005}.

\begin{figure}[htb]
\includegraphics[trim = 0mm 0mm 0mm 0mm,clip,width=8.6cm]
{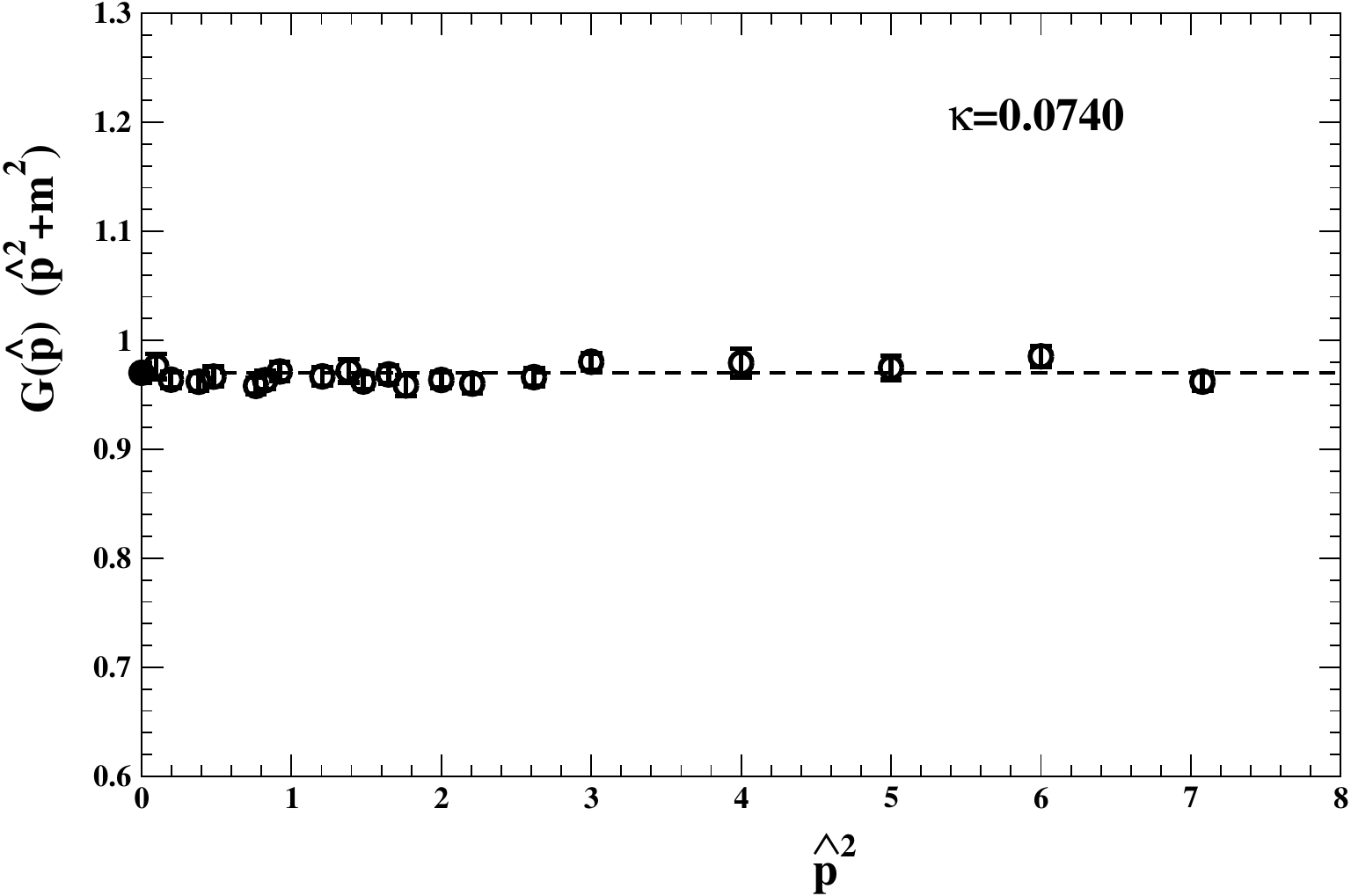}
\caption{The lattice data from
Ref.~\cite{Consoli:2020nwb} for the rescaled propagator in the
symmetric phase at $\kappa=0.074$, as a function of the lattice
momentum squared ${\hat p}^2$. The fitted mass from high ${\hat p}^2$,
viz.\ $m_{\rm{latt}}=0.2141\,(28)$, describes the data well, down
to ${\hat p}=0$.}
\label{0.074}
\end{figure}

\begin{figure}[htb]
\includegraphics[trim = 0mm 0mm 0mm 0mm,clip,width=8.6cm]
{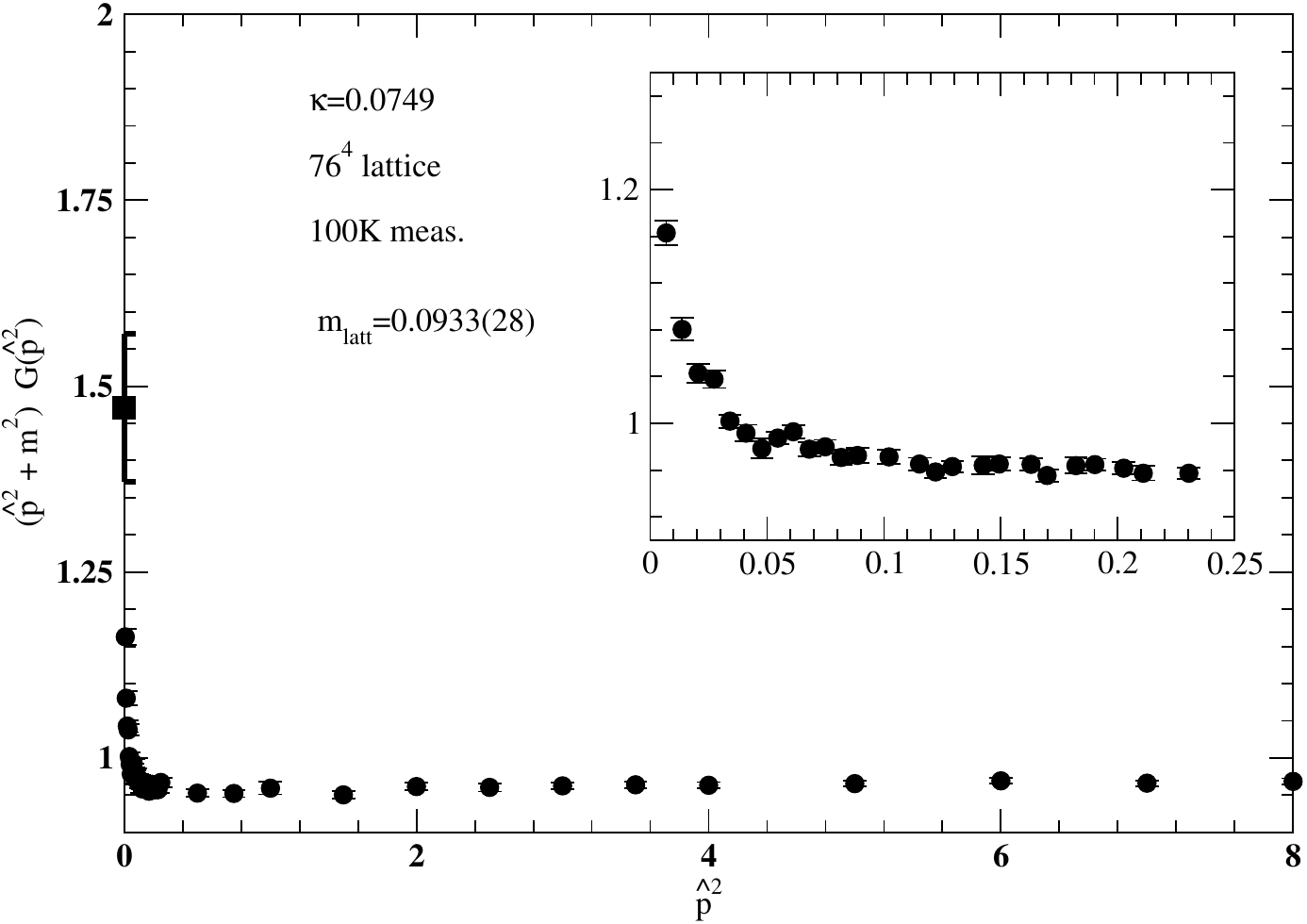}
\caption{The propagator data from
Ref.~\cite{Consoli:2020nwb}, for $\kappa=0.0749$, rescaled with the
lattice mass $M_H \equiv m_{\rm{latt}}=0.0933\,(28)$ obtained from the
fit to all data with ${\hat p}^2>0.1$. The peak at $p=0$ is
$M^2_H/m^2_h= 1.47\,(9)$, as computed from the fitted $M_H$ and the
zero-momentum mass $m_h=0.0769\,(8)$.}
\label{0933}
\end{figure}

\begin{figure}[htb]
\includegraphics[trim = 0mm 0mm 0mm 0mm,clip,width=8.6cm]
{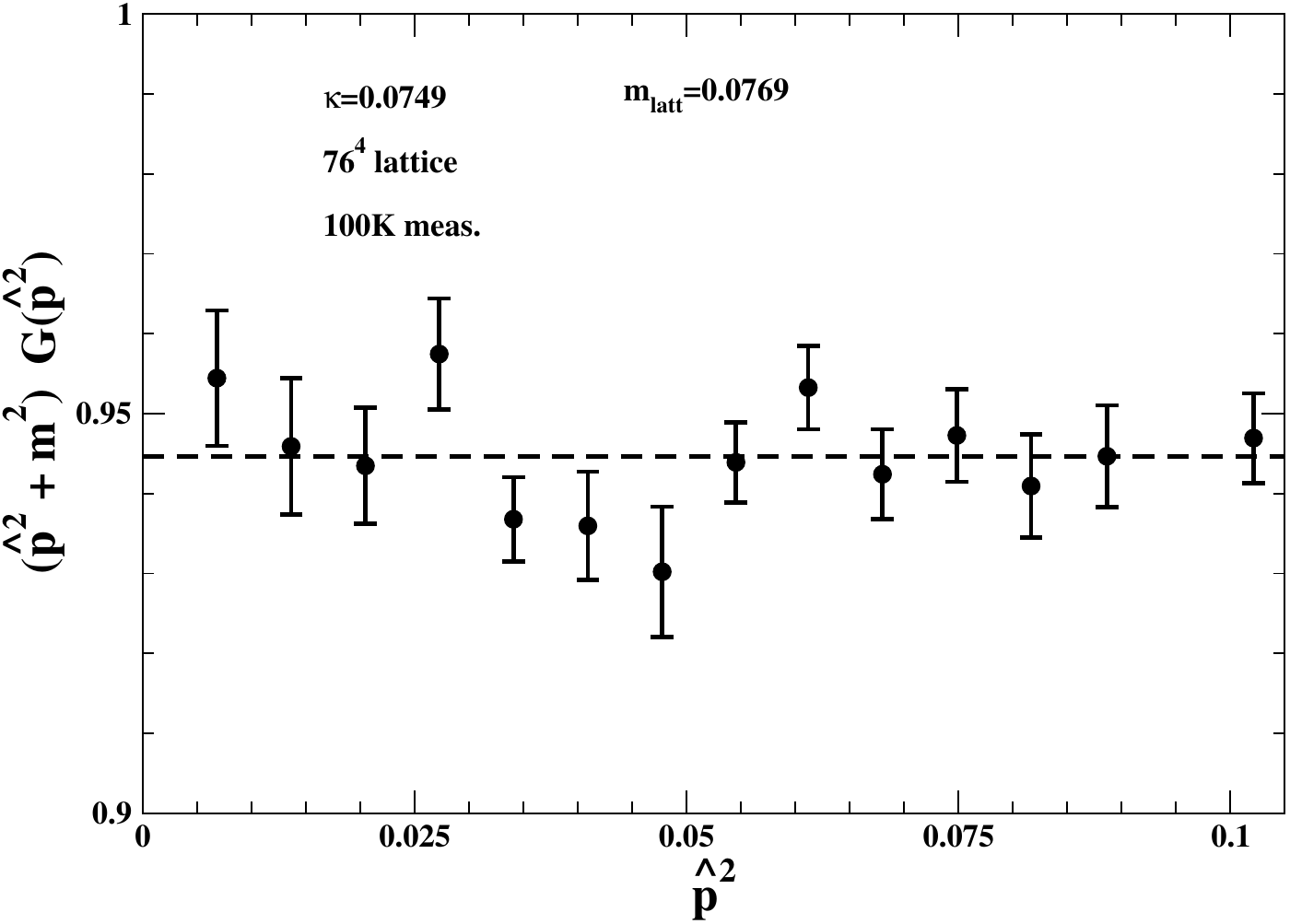}
\caption{The propagator data from
Ref.~\cite{Consoli:2020nwb} at $\kappa=0.0749$ for ${\hat p}^2<0.1$.
The lattice data were rescaled here with zero-momentum mass
$m_h=0.0769\,(8)$.}
\label{susce}
\end{figure}

The strategy adopted in Ref.~\cite{Consoli:2020nwb} was to first fit
the propagator data to the two-parameter form
\begin{equation}
\label{twoparameter}
G_{\rm{fit}}(p) \; = \;
\frac{Z_{\rm{prop}}}{{\hat p}^2+m^2_{\rm{latt}}}
\end{equation}
in terms of the squared lattice momentum ${\hat p}^2$. The data were then
rescaled by $({\hat p}^2+m^2_{\rm{latt}})$, so that deviations from a flat
plateau become immediately visible. While in the symmetric phase no
momentum dependence of the mass parameter was observed (see
Fig.~\ref{0.074}),
in the broken-symmetry phase there is a transition between two
regimes. As pointed out by Stevenson \cite{Stevenson2005}, by
rescaling all data with the mass from the higher-momentum fit, the
deviations from constancy become highly significant in the $p\to 0$
limit. In Ref.~\cite{Consoli:2020nwb} this was checked with a simulation
on a large $76^4$ lattice; see Figs.~\ref{0933} and \ref{susce}.

Crucially, extrapolation toward the
continuum limit with various lattices was consistent with the
expected scaling trend $M^2_H\sim L m^2_h$ from Eq.~(\ref{scale}).
Thus, from the lattice data for the propagator, one extracted
the numerical constant $c_2$ in the relation
\begin{equation}
\frac{M^2_H}{m^2_h}\Big|_{\rm latt}\sim L \, (c_2)^{-1} \; .
\end{equation}
The values of $(c_2)^{-1/2}=M_H\,m_h^{-1}\,L^{-1/2}$ are reported in
Table~\ref{latticetable} and, given their consistency, lead to a
final
\begin{table*}[t]
\caption{For various values of the hopping parameter $\kappa$ of
the 4D Ising model, we report the mass $M_H$ obtained from the
higher-momentum propagator data and the zero-momentum $m_h$ 
extracted from the lattice susceptibility $\chi$ through the relation
$(2\kappa\chi)^{-1/2}$; see Ref.~\cite{Consoli:2020nwb}. For
$\kappa=0.0749$ the three values of $M_H$ refer to high-momentum
fits for ${\hat p}^2>0.1$, ${\hat p}^2>0.15$ and ${\hat p}^2>0.2$
respectively on the $76^4$ lattice of ref.\cite{Consoli:2020nwb}.
The lattice cutoff $\Lambda_L\sim \pi/a$ and all masses are in units
of the inverse lattice spacing $a$. In the last column, we report
the combination $(c_2)^{-1/2}= M_H \, (m_h)^{-1} \, L^{-1/2}$.
The table is adapted from the original table in
Ref.~\cite{Consoli:2020nwb}.}
\begin{center}
\begin{tabular}{cccccc}
$\kappa$ & {\rm lattice} & $M_H$ & $ m_h$ & $[\ln(\Lambda_L/M_H)]^{-1/2}$ &
$(c_2)^{-1/2}$  \\
\hline \hline
     0.07512 & $32^4$ & 0.2062\,(41) & 0.1857\,(8) & 0.606\,(2) & 0.673\,(14)\\
\hline0.0751 & $48^4$ & $\sim0.200$ & 0.1796\,(5) & $\sim0.603$ & $\sim0.671$\\
\hline0.07504 & $32^4$ & 0.1723\,(34) & 0.1507\,(7)& 0.587\,(2) & 0.671\,(14)\\
\hline0.0749 & $76^4$ & 0.0933\,(28) & 0.0769\,(8) & 0.533\,(2) & 0.647\,(22)\\
\hline0.0749 & $76^4$ & 0.096\,(4)   & 0.0769\,(8) & 0.535\,(3) & 0.668\,(31)\\
\hline0.0749 & $76^4$ & 0.100\,(6)   & 0.0769\,(8) & 0.538\,(4) & 0.700\,(42)\\
\hline
\end{tabular}
\end{center}
\label{latticetable}
\end{table*}
determination
\begin{equation}
\frac{1}{\sqrt{c_2}}\; \sim \;0.67 \pm 0.03 \;.
\end{equation}
Therefore, in order to estimate $M_H$, the following strategy was adopted:
\begin{enumerate}[i)]
\item
First, one used the Gaussian-approximation relation
(valid in the whole range of $m_\Phi$ and not just for $m_\Phi=0$)
\begin{equation}
\label{gaussianmass}
M^2_H \; = \; \frac{\lambda}{3} \; ~ \; \phi_v^2 \; .
\end{equation}
\item
Second, the rescaling $Z_\phi=\phi_v^2/v^2=M^2_H/m^2_h$ was
extracted from the lattice data for the propagator, yielding
\begin{equation}
\phi^2_v \; \sim \; v^2 \, L \, (c_2)^{-1} \; .
\end{equation}
\item
Finally, in Eq.~(\ref{gaussianmass}) one used the leading-log
relation
\begin{equation}
\label{leading}
\lambda \; \sim \; \frac{16\pi^2}{3} \, L^{-1} \; .
\end{equation}
\end{enumerate}
This way, the constant $K$ was determined as
\begin{equation}
K \; = \; \frac{4\pi}{3} \, (c_2)^{-1/2} \; = \;
2.80 \pm 0.12 \; ,
\end{equation}
(or $K\sim 8\pi/9$), so that for $v\sim$
246~GeV one finds
\begin{equation}
\label{gaussianmass2}
(M_H)^{\rm Theor} \; = \; K v \;  \sim \; 690(30) \; {\rm GeV} .
\end{equation}

\subsection{From a one-component to an $N$-component theory}

Before considering $N$-component scalar fields, let us first summarise our
previous results for the one-component $\Phi^4$ theory:
\begin{enumerate}[a)]
\item
Lattice simulations in the Ising limit support a view of SSB as
a weak first-order phase transition. This is relevant because the Ising limit,
with a lattice coupling at the Landau pole, is known to saturate the triviality
bound in $\Phi^4$. Namely, for any non-zero renormalised coupling, it
represents the best approximation to the continuum limit.  \\
\item
This weak first-order scenario is also recovered in those known
approximations where the effective potential has the same basic
structure of a classical background plus a ZPE of
free-field-like fluctuations with some $\phi$-dependent mass. In
this sense, there is consistency with the basic ``triviality''
property of $\Phi^4$ theories in 4D. \\
\item
These known approximations to the effective potential predict
the same pattern of scales in Eq.~(\ref{scale}). As a consequence,
the mass parameter $M_H$, associated with the ZPEs that determine
the potential depth, is much larger than the mass parameter $m_h$
defined by the quadratic shape of the effective potential at the
minimum. The scalar propagator should thus deviate from a
standard one-mass structure. \\ 
\item
Lattice simulations of the propagator confirm the existence of these
deviations and are consistent with the
expected scaling trend $M^2_H\sim L m^2_h $ in Eq.~(\ref{scale}).  \\
\item
By combining analytical and numerical relations, the second
mass scale $M_H$ can be estimated to be about 700~GeV.
\end{enumerate}

Concerning now the validity of these results for a
four-component theory, as for the physical Higgs field, we first mention
Ref.~\cite{alles}, where, in an $O(N)$-invariant theory and within
the same Gaussian approximation to the effective potential
considered before, the weak first-order scenario is confirmed.
Indeed, SSB is recovered when in its symmetric phase the scalar
quanta are massless. Therefore, the phase transition occurs for
small but still positive values of the mass-squared parameter at
$\phi=0$. In addition, as anticipated, the basic
Eq.~(\ref{gaussianmass}) is also valid in the $O(N)$ theory after
replacing $\lambda \to \lambda/4!$ in the $\Phi^4$ interaction term.
It is necessary, however, to go into some more detail and show that,
as in the one-component case, also the $O(N)$ theory exhibits a similar
two-mass structure of the shifted-field propagator.

To this end, let us remind that in the Gaussian approximation, one
introduces a variational mass $\Omega$, for the field component with
non-zero vacuum expectation value, and a different mass parameter
$\omega$ for the unshifted fields. Minimisation of the Gaussian effective
potential then yields the following coupled equations \cite{alles,okopinska}:
\BE
\label{Omega1}
\Omega^2 \; = \; m^2_B + \frac{\lambda\phi^2}{2} + \frac{\lambda}{2}
I_0(\Omega) + \frac{\lambda}{6} (N-1) I_0(\omega) \; ,
\EE
\BE
\label{omega2}
\omega^2 \; = \; m^2_B + \frac{\lambda\phi^2}{6} + \frac{\lambda}{6}
I_0(\Omega) + \frac{\lambda}{6} (N+1) I_0(\omega) \; ,
\EE
\BE
\label{B3}
B \; \equiv \; m^2_B + \frac{\lambda\phi^2}{6} + \frac{\lambda}{2}
I_0(\Omega) + \frac{\lambda}{6}(N-1)I_0(\omega) \; = \; 0 \; .
\EE
with $I_0(M)$ being defined in Eq.(\ref{I_0}). We emphasise that the
$\omega$s are just variational parameters. Upon diagonalisation of the
$N \times N$ propagator matrix \cite{okopinska}, one finds $(N-1)$ fields
with inverse propagators ${\cal G}^{-1}(p)=p^2+B$, which become exactly
massless at the absolute minimum $\phi_v$, where $B=0$.

About the shifted field, the Gaussian mass is again
$\Omega^2(\phi_v)=\lambda\phi^2_v/3 \equiv M^2_H$ and the inverse
propagator has the same form as in Eq.~(\ref{propagatorGEA}), viz.\
\BE
\label{GEA_N}
G^{-1}(p) \; = \; p^2 +  M^2_H  A(p,\Omega,\omega) \; ,
\EE  where
now\footnote{The correct expression for the quadratic term
proportional to $(N+2)$ in the denominator of $A(p,\Omega,\omega)$ is
reported in Ref.~\cite{alles}.}
\begin{eqnarray}
\label{Ap_N}
\lefteqn{A(p,\Omega,\omega) \; = \;} \nonumber \\
&&\frac{ 1- J(p,\Omega) + \frac{J(p,\omega)}{3} - \frac{N+2}{9}
J(p,\Omega)J(p,\omega) } {1+ \frac{J(p,\Omega)}{2} +
\frac{N+1}{6}J(p,\omega)+\frac{N+2}{18} J(p,\Omega)J(p,\omega) } . 
\end{eqnarray}
Here, $J(p,M)$ is again as in Eq.~(\ref{JpM}). With the parameters
$\Omega=\Omega(\phi_v)$ and $\omega=\omega(\phi_v)$, which are
solutions of the minimisation in Eqs.(\ref{Omega1})--(\ref{B3}), we
can thus compute the screening factor $A(0,\Omega,\omega)$ that
fixes the proportionality relation between the Gaussian mass
$M^2_H=\Omega^2(\phi_v)$ and the zero-momentum limit $G^{-1}(p=0)
\equiv m^2_h$. In the one-component theory, the (two) coupled
minimisation equations amount to $J(0,M_H)= 1-\epsilon$ or
$A(0,M_H)\sim 2\epsilon/3 \sim L^{-1} $. Therefore,
confirming this result for the general $O(N)$ theory is equivalent
to showing that, to zeroth-order in $\epsilon$, i.e., by just
retaining terms of $\mathcal{O}(\epsilon L )\sim 1$, we find
$A^{(0)}(0,\Omega,\omega)=0$.

Then, by using the identity
\BE
\lambda[I_0(M)-I_0(0)] \; = \; -M^2 [J(0,M) + \epsilon]
\EE
and defining $X=J(0,\Omega)+\epsilon$, $Y= J(0,\omega)+\epsilon$,
the coupled minimisation equations yield
\BE
\label{coupledXY}
\frac{\Omega^2}{\omega^2}  =  \frac{3+Y}{X} \;\;\;\; \mbox{and} \;\;\;\; 
Y  =  \frac{9(1-X)}{(N+2)X-3} \; .
\EE
Furthermore, since
\BE
Y \; = \; X +\epsilon \ln(\Omega^2/\omega^2) \; = \; 
X + \epsilon \ln \left[ \frac{3+Y}{X} \right] \; ,
\EE
one can solve the two equations in
Eq.~(\ref{coupledXY}) iteratively, as $Y^{(0)}=X $, $Y^{(1)}=X +
\epsilon \ln [(3+Y^{(0)})/X]$, and so on. Therefore, to zeroth-order
in $\epsilon$, when $Y^{(0)}=X $ and $X$ becomes a solution of
\BE
(N+2)X^2+ 6X -9 \; = \; 0 \; ,
\EE
we find the sought result
\BE
\label{A0_N}
A^{(0)}(0,\Omega,\omega) \; = \; \frac{ 1-  \frac{ 2 } {3} X - \frac{ N+2 }
{9} X^2 } {1+ \frac{N+4} {6} X +\frac{N+2} {18} X^2} \; = \; 0 \; ,
\EE
implying
\BE
\label{A0omega}
A(0,\Omega,\omega) \; \sim \; \epsilon \; \sim \;  L^{-1} \; ,
\EE
so that
\BE
\label{mhN}
m^2_h \; \equiv \; M^2_H \, A(0,\Omega,\omega) \; \sim \; 
M^2_H \, L^{ -1} \; \ll \; M^2_H \; .
\EE
Finally, recovering the Gaussian mass in the higher-momen\-tum limit
where $A(p,\Omega,\omega) \lesssim 1$ proceeds along the same lines
as in the one-component theory.

Truly enough, despite having obtained the same type of
structure in Eqs.(\ref{mh1}) and (\ref{mhN}), one may object that we
have not carried out lattice simulations of the propagator in the
$O(4)$ case. Thus one could still wonder about the coefficient $c_2$
of the logarithmic slope, which is crucial for the prediction of
$M_H$ in Eq.~(\ref{gaussianmass2}). Since the effective potential is rotationally invariant, 
it is conceivable that basic properties of its shape, such as the relation between 
the second derivative at the minimum and its depth, should be the same as 
in a one-component theory.  For a quantitative argument, however, one can reason as follows. From the
relations in Eq.~(\ref{scale}), one finds $m_h \ll M_H$ for very large
$\Lambda$. But $M_H$ is independent of $\Lambda$, so that by decreasing
$\Lambda$ and so increasing the lower mass, the latter would naturally
approach its maximun value $(m_h)^{\rm max}\sim M_H$ when the cutoff
$\Lambda$ becomes a few times $M_H$. Therefore, in our approach, the numerical
estimate Eq.~(\ref{gaussianmass2}) could also be used to place an upper bound
on the cutoff-dependent $m_h$, i.e.,
\begin{equation}
\label{upperbound}
(m_h)^{\rm max} \sim (M_H)^{\rm Theor}  \sim
690(30) \; {\rm GeV} .
\end{equation}
We can thus check the consistency of this prediction from the
one-component theory and compare relation (\ref{upperbound}), with
the existing theoretical upper bounds from lattice simulations of
the $O(4)$ theory. In this case, we find a very good agreement with
Lang's \cite{lang} and Heller's \cite{heller} values, viz.\
$(m_h)^{\rm max}=670\,(80)$~GeV and $(m_h)^{\rm max}=710\,(60)$~GeV,
respectively, depending on slightly different assumptions about the
magnitude of the minimum ultraviolet cutoff. Actually, by combining
these two estimates, the resulting value $(m_h)^{\rm max} \sim
690\,(50)$~GeV would coincide exactly with our expectation
(\ref{upperbound}) deduced from Eq.~(\ref{gaussianmass2}). In this
sense, our arguments can also be reversed. By assuming these two old
theoretical upper bounds, we could have predicted the value of $M_H$
without performing our own lattice simulations of the propagator. At
the same time, the important theoretical role of $(m_h)^{\rm max}$
should not make us forget that in the real world $m_h=125$~GeV.
Therefore, if there is a second resonance with $M_H\sim 690$~GeV,
the ultraviolet cutoff $\Lambda$ should be extremely large.

\section{Basic phenomenology of the second resonance}

By assuming the description of SSB in $\Phi^4$ theories given in
Sec.~2, we will now explore the implications for the full Standard
Model. We first observe that, with a large value $M_H \sim$~690 GeV,
including the ZPEs of all known gauge and fermion fields at the
Fermi scale would represent a small radiative
correction.\footnote{By subtracting quadratic divergences or using
dimensional regularisation, the logarithmically divergent terms in
the ZPEs of the various fields are proportional to the fourth power
of the mass. Thus, in units of the pure scalar contribution, one finds $ (6
M^4_w + 3 M^4_Z)/M^4_H \lesssim 0.002$ and $12 m^4_t/M^4_H\lesssim
0.05$.} As in the early days of the Standard Model, one could thus
adopt the perspective of explaining SSB within the pure scalar
sector and restrict the analysis to a region around the Fermi scale
that is not much larger than a few TeV. Then, once $\lambda^{\rm
(p)}(v)$ and $\lambda(v)$ have the same value as in
Eq.~(\ref{same}), the different evolution of the two couplings at
asymptotically large energies should remain unobservable. Checking
our proposed mechanism for SSB then requires the observation of a
second resonance and analysing its phenomenology.

Before entering more phenomenological aspects, let us first go back
to the general form of the Euclidean propagator in
Eqs.(\ref{propagatorGEA}) and (\ref{GEA_N}), say \BE
\label{GEA_general} G^{-1}(p)= p^2 +  M^2_H A(p) \EE
This structure corresponds to $G^{-1}_h(p)\sim  p^2 + m^2_h$ at low $p^2$, where $A(p)\sim \epsilon$, and to  
$G^{-1}_H(p)\sim p^2 + M^2_H$ at larger $p^2$, where $A(p) \sim 1$, in agreement with lattice simulations.  Continuation to Lorentzian space, will then produce corresponding real parts
$Re[G^{-1}_h(s)]\sim -s + m^2_h$ and $Re[G^{-1}_H(s)]\sim -s +
M^2_H$ defining vastly different mass-shell
regions $m_h$ and $M_H$  \footnote{Note that a large Euclidean momentum $p^2$
can translate to a massless on-shell Lorentzian particle with ${\bf
p}^2 - (p_0)^2=0$ or, more generally, to mass-shell regions much smaller than $p^2$, see the discussion in
\cite{donoghue}.}. These correspond to two types of ``quasi-particles'', almost
freely propagating in the broken-symmetry vacuum, in analogy with
phonons and rotons \footnote{This analogy is most natural in a
description of SSB as a (weak) first-order phase transition where
the broken-symmetry vacuum can be thought as originating from the
condensation of the quanta of the symmetric phase (with physical
mass $m^2_\Phi >0$). These elementary quanta would then be the
analogs of the He-4 atoms while the two quasi-particles with mass
$m_h$ and $M_h$ would be the analogs of the two collective
excitations of the system (delocalized density fluctuations at long
wavelengths and localized vortex modes at shorter distances) whose hybridization \cite{gov} provide a good
description of the energy spectrum of the superfluid.} in
superfluid He-4 .

As for the relevant phenomenology, a Higgs resonance with mass
$M_H\sim$ 700 GeV is usually believed to be a broad resonance due to
strong interactions in the scalar sectors. This belief derives from
two arguments, namely the definition of $M_H $ from the quadratic
shape of the potential, which is not valid in our case, and the
tree-level calculation of longitudinal $WW$ scattering. Then, at
high energies, due to an incomplete cancelation of graphs, the mass
in the scalar propagator is effectively promoted to a coupling
constant. For the sake of clarity, we shall consider this second
argument and make use of the tree-level expression reported by Veltman and
Yndurain \cite{yndurain}
\begin{equation}
A_{\rm ww} \; = \; \frac{g^2}{4 M^2_w} \left[a_{\rm tree}(s) +
a_{\rm tree}(t)+ a_{\rm tree}(u)\right]  \; ,
\end{equation}
where $g=g_{\rm gauge}$, $M^2_w=g^2v^2/4$,
and the tree-level amplitude is
\begin{equation}
a_{\rm tree}(x) \; = \; x + \frac{x^2}{M^2_H -x}  \; .
\end{equation}
Therefore, for $|x| \gg M^2_H$ as in
a multi-TeV collider, the tree-level amplitude is governed by a
contact coupling
\begin{equation}
\lambda_0 \; = \; \frac{3 M^2_H}{v^2}
\end{equation}
as in a pure $\Phi^4$ theory. Notice that the tree-level calculation
yields $\lambda_0$, while, from the $\Phi^4$ effective potential, we
found
\begin{equation}
\lambda(v) \; = \; \frac{3M^2_H}{\phi^2_v} \; = \;
\frac{3m^2_h}{v^2} \; = \; \frac{m^2_h}{M^2_H} \, \lambda_0 \; ,
\end{equation}
which derives from the assumed ``triviality'' of the theory. To
understand the replacement, let us recall the precise formulation of
the Equivalence Theorem given by Bagger and Schmidt \cite{bagger}.
This is a non-perturbative statement in the sense that it holds to
all orders in the scalar self-interactions, up to $\mathcal{O}(g^2)$
corrections. One then expects that resumming all higher-order graphs
in longitudinal $WW$ scattering gives the same result as in a pure
$\Phi^4$ theory, if Goldstone $\chi\chi$ diagrams are resummed with
the $\beta-$function of $\Phi^4$. This resummation, for $M_H\sim
700$~GeV, means replacing $\lambda_0 \sim 24$ with $\lambda(v) \sim
0.78$. For this reason, no large effect proportional to $\lambda_0$
should be visible at the Fermi scale or at the relatively nearby
energies of a multi-TeV collider (where both $\lambda(E) $ and
$\lambda^{\rm (p)}(E)$ differ from their common value 0.78 by
negligible terms). In this sense, the second resonance will mimic a
conventional Higgs particle of mass $M_H$, provided the
cutoff-independent ratios $M^2_H/v$ and $(M_H/v)^2$, in
the three-point and four-point scalar couplings, respectively, are
rescaled as follows \cite{Castorina}:
\begin{equation}
\frac{M^2_H}{2 v} \; \to \; \epsilon_1 \, \frac{M^2_H}{2 v}\; , \;\;\;
\frac{M^2_H}{8v^2} \; \to \; \epsilon_2 \, \frac {M^2_H}{8 v^2} \; ,
\end{equation}
with
\begin{equation}
\epsilon^2_1 \; = \; \epsilon_2\equiv \frac{1}{Z_\phi} \; = \;
\frac{m^2_h}{M^2_H} \; .
\end{equation}
We can thus predict $\Gamma(H\to WW)$
and $\Gamma(H\to ZZ)$ from their conventional values by replacing
the large width $\Gamma^{\rm conv}(H \to WW+ZZ) \sim G_F M^3_H$ with
the corresponding value $\Gamma(H \to WW+ZZ) \sim M_H (G_F m^2_h)$,
which retains the same phase-space factor $M_H$, but has a coupling
rescaled by the small ratio $m^2_h/M^2_H\sim  0.032$. In a first
estimate, just to display the large renormalisation effect, we can
retain the lowest-order values \cite{widths} and (for $M_H \sim
700$~GeV) we find
\begin{eqnarray}
\label{rel1}
\Gamma(H \to ZZ) & \sim & \frac{M_H}{700 \; {\rm GeV}} \,
\frac{m^2_h}{(700 \; {\rm GeV})^2} \, 50.1 \; {\rm GeV} \; \sim \nonumber \\
&& \frac{M_H}{700\;{\rm GeV}} \, 1.6 \; {\rm GeV}
\end{eqnarray}
and
\begin{eqnarray}
\label{rel2}
\Gamma(H \to WW) & \sim & \frac{M_H}{700~{\rm GeV}} \,
\frac{m^2_h}{(700~{\rm GeV})^2} \, 102.6~{\rm GeV} \; \sim \nonumber \\
&& \frac{M_H}{700\;{\rm GeV}} \, 3.3 \; {\rm GeV} \; .
\end{eqnarray}
On the other hand, the decays into fermions, gluons, photons, \ldots should
be unchanged,\footnote{A possible exception concerns the decay width
$\Gamma(H\to\gamma\gamma)$. For the precise value $M_H= 700$ GeV,
the full estimate of Ref.~\cite{handbook} is $\Gamma(H \to
\gamma\gamma)\sim 29$~keV. But this estimate contains the
non-decoupling (also called ``$-2$'') term proportional
to $G_F M^3_H$, whose existence (or not) in the $WW$ contribution has
been discussed at length in the literature. At present, the general
consensus is that this term has to be there, the only exceptions
being the unitary-gauge calculations of Gastmans, Wu, and Wu
\cite{GWuWu,WuWu,WuWu2}, and the dispersion-relation approach of
Christova and Todorov \cite{Todorov}. We believe that, in the
context of a second Higgs resonance, which does {\it not} \/couple
to longitudinal $W$s proportionally to its mass but with the same
typical strength as the low-mass state at 125 GeV, the whole issue
could be reconsidered (especially if one realises how delicate the
matter actually is \cite{Melnikov}).  The point is that, for the
range of masses around $M_H= 700$~GeV, there are strong cancelations
between the $WW$ and $t \bar t$ contributions. As we have checked,
dropping the non-decoupling term, or replacing it with $M_H (G_F m^2_h)$,
could easily change the lowest-order value (i.e., without QCD corrections
in the top-quark graphs) by an order of magnitude. For this reason, a
partial decay width $\Gamma(H \to\gamma\gamma)=\mathcal{O}(100)$~keV
cannot be excluded. In any case, this issue, while conceptually
relevant, is unimportant for a first estimate of the total $H$
width.}
yielding
\begin{equation}
\begin{array}{c}
    \Gamma(H\to {\rm fermions+ gluons+ photons+\ldots}) \; \sim \\[1mm]
\displaystyle\frac{M_H}{700~{\rm GeV}} \, 27 \; {\rm GeV} \; .
\end{array}
\end{equation}
Therefore, for $M_H=660\div720$~GeV, one would expect a total width
$\Gamma_H\equiv \Gamma( H \to all)= 30\div 33$~GeV.

However, the previous estimates do not include the new
contributions from the decays of the heavier $H$ into a pair of the
lower state $h$ at 125~GeV. Addressing this point requires first to
consider the problem of $h$-$H$ mixing, which could slightly change
the normalisation of the two states. This problem should also be considered
in the framework of a ``trivial'' theory, such as the pure scalar sector,
where, far from the Landau pole, all large interaction effects are
reabsorbed into the mass parameters and in the vacuum structure. This is
well illustrated in the previous propagator structure implied
by the Gaussian Effective Action, which by definition is consistent with
``triviality'', and is also confirmed by the fits to the lattice data
\cite{Consoli:2020nwb},  where the residuals $Z_{\rm{prop}}=0.94\div 0.96$
were slightly smaller than unity, in the respective momentum regions, as
expected for almost free, quasi-particle states. For this reason, residual
mixing effects are not expected to play a significant role in the on-shell
normalisation of the two states. But, of course, in the full Standard Model,
there will also be additional contributions to $h$-$H$ mixing through the
gauge and fermion loops. We will present in Sec.~5 a coupled-channel 
formalism that, in our opinion, provides a convenient framework to address
this problem. However, carrying out a full calculation goes beyond our
present scope. Still, we can try to further improve on our analysis. 
For the state at 125~GeV, its on-shell normalisation is constrained by LHC
measurements of its couplings that agree with the SM predictions at the
5$\div 10$\% level; see e.g.\ Figs.~7 and 8 of Ref.~\cite{ortona}. 
Therefore, if we assume that an observable $h$-$H$ mixing effect in the
on-shell normalisation of the $H$ resonance has a similar order of magnitude
(so that it can be neglected at the present level of accuracy), we get two
definite predictions. First, by just using rotational $O(4)$ invariance, we
can relate the coupling for the main decay width $\Gamma(H\to hh)$ to the
corresponding coupling for a decay into a pair of longitudinal $Z$s. Up to a
small phase-space correction $f(m_h)/f(M_z)\sim 0.97$, with
$f(m)=\sqrt{1-4(m/M_H)^2}$, this would give an additional contribution
$\Gamma(H\to hh) \sim 1.52\,(7)$ GeV and a branching ratio
$B(H\to hh) \sim 0.046$. As we will discuss in Sec.~4, this estimate is well
consistent with the experimental upper bounds on $B(H\to hh)$ obtained
from the $b \bar b + \gamma\gamma$ channel, thus implying
$\Gamma_H < 35~{\rm GeV}$.  Secondly, as in Ref.~\cite{signals}, assuming a
standard on-shell normalisation of the $H$ state and thus neglecting
$h$-$H$ mixing altogether, there is a very precise test in the ``golden''
four-lepton channel that does {\it not} \/require knowledge of the total
width. Indeed, it just relies on two assumptions:
\begin{enumerate}[a)]
\item
A resonant four-lepton production through the chain $H \to ZZ \to 4l$
~ ($l=e,\mu$);
\item
Exactly the same estimate of $\Gamma(H \to ZZ)$ from Eq.~(\ref{rel1}).
\end{enumerate}
Therefore, by defining $\gamma_H=\Gamma_H/M_H$, we find a fraction
\begin{equation}
\begin{array}{l} \displaystyle
B( H \to ZZ) \; = \; \frac{\Gamma( H \to ZZ)}{\Gamma_H} \; \sim \;
\frac{1}{\gamma_H} \, \frac{50.1}{700}  \,
\frac{m^2_h}{(700\;{\rm GeV})^2}
\end{array}
\end{equation}
that will be replaced in the cross section approximated
by the on-shell branching ratios
\begin{eqnarray}
\label{exp3}
\sigma_R (pp\to H\to 4l) & \sim & \sigma (pp\to H) \, B( H \to ZZ)
\times \nonumber \\[1mm]
&& 4 B^2(Z \to l^+l^-) \; .
\end{eqnarray}
This should be a good approximation for a relatively narrow
resonance, so that one predicts a precise correlation
\begin{eqnarray}
\label{exp33}
\gamma_H \, \sigma_R (pp\to H\to 4l) & \sim &
\sigma(pp\to H) \, \frac{50.1}{700} \, \frac{m^2_h}{(700~{\rm GeV})^2}
\times \nonumber \\
&& 4 B^2(Z \to l^+l^-) \; ,
\end{eqnarray}
which can be compared to the LHC data.

Since $4B^2( Z \to l^+l^-)\sim 0.0045$, in order to check our prediction
the last needed ingredient is the total production cross section
$\sigma (pp\to H)$, which in our case will mainly result from the
gluon-gluon Fusion (ggF) process. In fact, the other production mechanism
through Vector-Boson Fusion (VBF) plays no role here, since the large
coupling to longitudinal $W$s and $Z$s is suppressed by the small
coefficient $m^2_h/M^2_H\sim  0.032$. As a consequence, the
traditionally large VBF cross section
$\sigma^{\rm VBF}(pp\to H)\sim  300$~fb is reduced to about 10~fb and can be
safely neglected in comparison with the pure ggF contributions of
$\mathcal{O}(10^3)$~fb. Indeed, for $pp$ collisions at 13~TeV and with a
typical $\pm 15\%$ uncertainty (due to the parton distributions, the choice
of $\mu$ in $\alpha_s(\mu)$, and other effects), we will adopt the value
\cite{yellow} $\sigma^{\rm ggF} (pp\to H)=1090\:(170)$~fb, which also
accounts for the possible mass range $M_H=660\div 700$ GeV.

In conclusion, for $m_h=$ 125 GeV, one obtains a
prediction that, for a not too large $\gamma_H$ where Eq.~(\ref{exp3})
starts to lose validity, is formally insensitive to the value of $\Gamma_H$
and can be directly compared to the four-lepton data
\begin{equation}
\label{exp34}
\left[\gamma_H \, \sigma_R (pp\to H\to 4l)\right]^{\rm Theor}
\; \sim \; (0.011 \pm 0.002)~ {\rm fb} \; .
\end{equation}

\section{Some experimental signals from LHC}

To test our definite prediction
$(M_H)^{\rm Theor}=690\pm10\,(\mbox{stat})\pm20\,(\mbox{sys})$~GeV,
one should look for deviations from
the background nearby. This means that local deviations should not
be downgraded by the so called ``look elsewhere'' effect. At the
same time, given the present energy and luminosity of LHC, the
second resonance, if there, is too heavy to be seen unambiguously in
all possible channels. In this sense, one should remember the $h(125)$
discovery, which, at the beginning, was producing no signals in the
important $b\bar b$ and $\tau^+ \tau^-$ channels.

After this premise, given the expected large branching ratio
$B(H\to t\bar t)=(75\div 80)\%$, the most natural place to look for
the new resonance would be in the $t\bar t$ channel. However, in
the relevant region of invariant mass $m(t\bar t)=620\div820$~GeV,
CMS measurements \cite{CMS_top} give a background cross section
$\sigma(pp\to t\bar t)=107\pm7.6$~pb which is about 100 times
larger than the expected signal
$\sigma(pp\to H \to t \bar t) \lesssim 1$~pb \footnote{Still, very recent
ATLAS measurements in the $t \bar t$ channel show some excess around 675 GeV
(see Fig.~12 page 34 of Ref.~ \cite{ATLAS_conf}), especially for those events
where the tracks of the two final leptons are at large angles.}
\footnote{Interestingly,
the process $pp\to t\bar t t\bar t$ has now been observed by ATLAS
\cite{ATLAS4TOP} and CMS \cite{CMS4TOP}.
Both experiments find cross sections that are
somewhat larger than the SM estimate $\sigma_B(pp\to t \bar t t \bar
t)=12.0 \pm 2.4$ fb. Namely, $22.5^{+6.6}_{-5.5}$ fb (ATLAS) and
$17.7^{+4.4}_{-4.0}$ fb (CMS). This excess could indicate the
process $pp \to H$ with the $H$ resonance decaying into a virtual
pair of $h=h(125)$ followed by two $h \to t \bar t$ decays. For this
reason, it would be interesting to determine the invariant mass
distribution of the $t \bar t t \bar t$ system.}. For this reason,
in Refs.~\cite{signals,ichep,CFF,universe} the phenomenological
analysis was focused on available channels with relatively smaller
background, namely:
\begin{enumerate}[i)]
\item
ATLAS  ggF-like four-lepton events;
\item
ATLAS high-mass inclusive $\gamma\gamma$ events;
\item
ATLAS and CMS $(b\bar b + \gamma\gamma)$ events;
\item
CMS $\gamma\gamma$ events produced in $pp$
diffractive scattering.
\end{enumerate}

\subsection{The ATLAS  ggF-like 4-lepton events}

As a first sample, let us focus on the ATLAS charged four-lepton
channel \cite{atlas4lHEPData}. In the search for a heavy scalar
resonance $H$, decaying through the chain $H\to ZZ \to 4l$, the
ATLAS experiment has performed a sophisticated analysis in which the
four-lepton events, depending on their topology, are divided into
ggF-like and VBF-like events. By expecting our second resonance to
be produced through gluon-gluon fusion, we have considered the
ggF-like category, which, depending on the degree of contamination
with the background, is further divided into four mutually exclusive
subcategories: ggF-high-4$\mu$, ggF-high-2e2$\mu$, ggF-high-4e,
ggF-low.

The only sample which is homogeneous from the point of view of the
selection and has a sufficient statistics is the ggF-low category
which contains a mixture of all three final states.
We understand  that the ggF-low sample is certainly less pure as
compared to the ggF-high samples, and it is true that it includes
the dominating contribution from other sources of non-resonant $ZZ$
events. On the other hand, this background was, according to ATLAS,
carefully evaluated with a total quoted (statistical + systematical)
uncertainty of less than 5\% in the relevant region. As such, there
is no reason not to consider it as our best estimate and safely compare
background and observed events. Since in the region of
invariant mass around 700~GeV the energy resolution of these events
varies considerably,\footnote{The resolution varies
from about 12~GeV for $4e$ events, to 19~GeV for $2e2\mu$, and up to
24~GeV for $4\mu$.}
it is natural to adopt a large-bin visualization to avoid spurious
fluctuations between adjacent bins. The numbers of events for this category
are shown in Table~\ref{leptontable}, together with the background
estimated by ATLAS \cite{atlas4lHEPData}.

\begin{table*}
\caption{For a luminosity of 139 fb$^{-1}$, we report the observed
ATLAS \cite{atlas4lHEPData} ggF-low events $\rm{N}_{\rm EXP}(E)$ and
the corresponding estimated background $\rm{N}_{\rm B}(E)$ in
the range of invariant mass $m(4l)=E=530\div 830$~GeV. In view of
the considerable difference in the energy resolution of the various
types of four-lepton events, to avoid spurious migrations between
neighbouring bins, we have grouped the data into larger bins of
60~GeV, centred at 560, 620, 680, 740, and 800~GeV. These correspond to
the 10 bins of 30~GeV, from $545\,(15)$~GeV to $815\,(15)$~GeV; see
Ref.~\cite{atlas4lHEPData}. In this energy range, the uncertainties
in the background are below 5\% and have been neglected. The
statistical errors of $\rm{N}_{\rm EXP}(E)$ are not reported by
ATLAS and will be assumed to be given by $\sqrt {\rm{N}_{\rm EXP}}$,
as for a Poisson distribution.}
\begin{center}
\begin{tabular}{cccc}
$\rm E$ [GeV] & $\rm{N}_{\rm EXP}(E)$  & $\rm{N}_{\rm B}(E)$ &
$\rm{N}_{\rm EXP}(E) - \rm{N}_{\rm B}(E)$ \\
\hline \hline
$560\,(30)$ & 38$\pm 6.16$ & 32.0 & $ 6.00 \pm 6.16$  \\
\hline
$620\,(30)$ & 25$\pm 5.00$ & 20.0 & $5.00 \pm 5.00$  \\
\hline
$680\,(30)$ & 26$\pm 5.10$ & 13.04 & $12.96 \pm 5.10$ \\
\hline
$740\,(30)$ & 3$\pm 1.73$ & 8.71& $-5.71 \pm 1.73$  \\
\hline
$800\,(30)$ & 7$\pm 2.64$ & 5.97 & $1.03 \pm 2.64$  \\
\hline
\end{tabular}
\end{center}
\label{leptontable}
\end{table*}

From this comparison, one finds a considerable excess over
the background, in the bin centred around 680~GeV, followed by a
sizeable defect in the next bin centred around 740~GeV. The
simplest explanation for these two simultaneous features would be
the existence of a resonance of mass $M_H\sim 700$~GeV which, above
the Breit-Wigner peak, produces the characteristic
negative-interference pattern proportional to $(M^2_H-s)$.

To check this idea, we will exploit the basic model where the $ZZ$
pairs, each of which subsequently decaying into a charged $l^+l^-$
pair, are produced by various mechanisms at the parton level.
Depending on the invariant mass of the four final leptons $E\equiv
m(4l)$, this gives rise to a smooth distribution of background events
$N_b(E)$, proportional to a background cross section $\sigma_b(E)$.

To describe the effects of a resonance, let us denote by $A_b(E)$
the background amplitude, whose squared modulus is proportional to
$\sigma_b(E)$, and by $A_R(E)$ the amplitude describing $ZZ$
production through the intermediate resonance $H$. We thus obtain a
total amplitude $A_T(E)= A_b(E) + A_R(E)$, whose square modulus will
be proportional to the total cross section. Now, by defining $s=E^2$
and introducing the complex resonance mass $M^2_R=M^2_H-iM_H\Gamma_H$
for a relatively narrow resonance, the resonant
amplitude can be approximated in terms of some constant $a_R$ as
\BE
A_R(E) \; \sim \; \frac{a_R}{s- M^2_R } \; .
\EE
Introducing then a (common) phase-space normalisation constant
$\xi$ such that $\xi|A_b(E)|^2 = \sigma_b(E)$ and
$\xi|A_R(E)|^2=$ $\sigma_R(E)$, with
\BE
\sigma_R(E) \; = \; \frac{\xi a^2_R }{(s-M^2_H)^2+(\Gamma_H M_H)^2} \; ,
\EE
the total cross section $\sigma_T(E)= \xi |A_T(E)|^2$ can be
conveniently expressed as
\begin{eqnarray}
\sigma_T(E) &  = & \sigma_b(E) +
\frac{2(M^2_H-s)\,\Gamma_H M_H}{(s-M^2_H)^2+(\Gamma_H M_H)^2} \,
\sqrt{\sigma_R\sigma_b(E)} + \nonumber \\[0.5mm]
&&\frac{(\Gamma_H M_H)^2 }{(s-M^2_H)^2+(\Gamma_H M_H)^2}\,\sigma_R \; ,
\label{sigmat}
\end{eqnarray}
where we have introduced the resonance peak cross section at $s=M^2_H$
defined as $\sigma_R= \xi a^2_R/ (M^2\Gamma^2_H) $ and assumed positive
interference below the peak as suggested by the data.

\begin{figure}[ht]
\includegraphics[width=0.4\textwidth,clip]{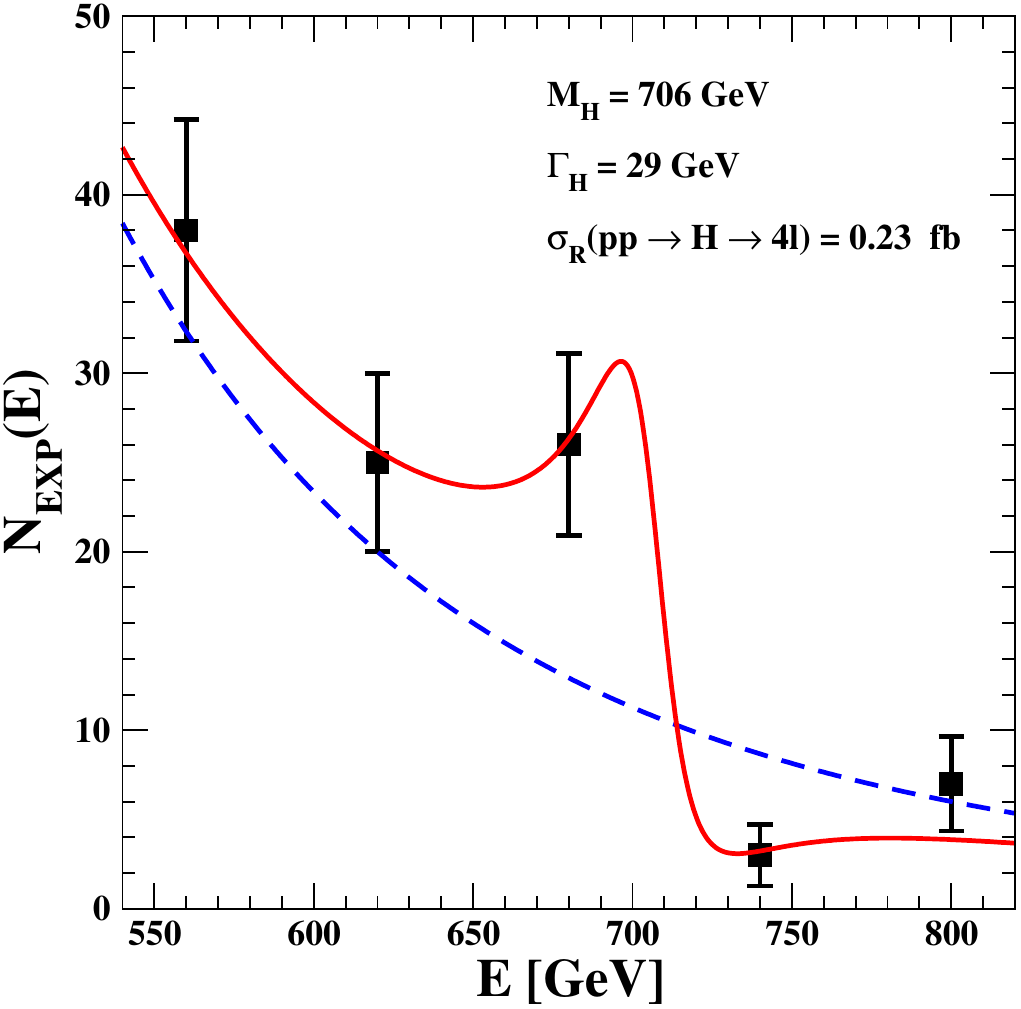}
\caption{The values $\rm{N}_{\rm EXP}(E)$ in Table~\ref{leptontable} for
ATLAS \cite{atlas4lHEPData} data vs.\ the corresponding
$\rm{N}_{\rm TH}(E)$ in Eq.~(\ref{NTH}) (solid red curve). The resonance
parameters are $M_H=706$~GeV, $\gamma_H=0.041$, $\sigma_R=0.23$~fb and the
ATLAS background (dashed blue curve) is approximated as
$N_b(E)= A\times({\rm 710~GeV}/E)^{\nu}$, with $A=10.55$ and $\nu= 4.72$.}
\label{4lepton}
\end{figure}

Now, an accurate description of the ATLAS background can be obtained
in terms of a power law $N_b(E)\sim A\times({\rm 710~GeV}/E)^{\nu}$,
with $A\sim 10.55$ and $\nu\sim 4.72$. Then, by simple
redefinitions, the theoretical number of events can be expressed as
\BE
N_{TH}(E) \;=\; N_b(E)+\frac{P^2+2P\,x(E)\,\sqrt{N_b(E)}}{\gamma^2_H+x^2(E)}\;,
\label{NTH}
\EE
where $x(E)=(M^2_H -E^2)/M^2_H$, $P \equiv \gamma_H\sqrt{N_R} $,
and $N_R=\sigma_R\times {\cal A } \times 139$~fb$^{-1}$ denotes the extra
events at the resonance peak, for an acceptance ${\cal A }$.

As for the acceptance, one can adopt a value ${\cal A }\sim 0.38$ by
averaging the two extremes, viz.\ 0.30 and 0.46, for the ggF-like category
of events \cite{atlas4lHEPData}. As a consequence, the resonance
parameters are affected by an additional uncertainty. Nevertheless,
to have a first check, in Refs.~\cite{CFF,universe} the experimental
number of events given in Table~\ref{leptontable} was fitted with
Eq.~(\ref{NTH}). The results were: $M_H=706\,(25)$~GeV,
$\gamma_H=0.041\pm 0.029$ (corresponding to a total width
$\Gamma_H=29\pm 20$~GeV), and $P=0.14 \pm 0.07$. From these numbers one
obtains $N_R\sim 12^{+15}_{-9}$ and $\sigma_R\sim 0.23^{+0.28}_{-0.17}$~fb.
The theoretical values are shown in Table~\ref{leptontable2} and a
graphical comparison in Fig.~\ref{4lepton}.

\begin{table*}
\caption{The observed ATLAS \cite{atlas4lHEPData} ggF-low events and our
theoretical prediction in Eq.~(\ref{NTH}), for $M_H=706$~GeV,
$\gamma_H= 0.041$, $P = 0.14$.}
\begin{center}
\begin{tabular}{cccc}
$\rm E$ [GeV] & $\rm{N}_{\rm EXP}(E)$  & $\rm{N}_{\rm TH}(E)$ & $\chi^2$ \\
\hline \hline
$560\,(30)$ & 38$\pm 6.16$ & 36.72 & 0.04  \\
\hline
$620\,(30)$ & 25$\pm 5$ & 25.66 & 0.02  \\
\hline
$680\,(30)$ & 26$\pm 5.10$ & 26.32 & 0.00 \\
\hline
$740\,(30)$ & 3$\pm 1.73$ & 3.23& 0.02  \\
\hline
$800\,(30)$ & 7$\pm 2.64$ & 3.87 & 1.40  \\
\hline
\end{tabular}
\end{center}
\label{leptontable2}
\end{table*}

The quality of the fit is good, but error bars are large and the test
of our picture is not very stringent. Still, with the partial width from
Sec.~3, viz.\ $\Gamma(H\to ZZ)\sim 1.6$~GeV, and fixing $\Gamma_H$ to its
central value of 29~GeV, we find a branching ratio $B(H\to ZZ)\sim0.055$ that,
for the central value $\sigma^{\rm ggF} (pp\to H)\sim  923$~fb from
Ref.~\cite{yellow} at $M_H=$ 700 GeV, would imply a theoretical peak cross
section $(\sigma_R)^{\rm theor}=923 \times 0.055 \times 0.0045 \sim 0.23$~fb,
which coincides with the central value from our fit. Moreover, from the central
values $\langle\sigma_R\rangle=0.23$~fb and $\langle\gamma_H\rangle=0.041$,
we find $\langle\sigma_R\rangle \times \langle\gamma_H\rangle\sim0.0093$~fb,
in accordance with Eq.~(\ref{exp34}).

\begin{figure}[ht]
\centering
\includegraphics[trim = 0mm 0mm 0mm 0mm,clip,width=8.6cm]
{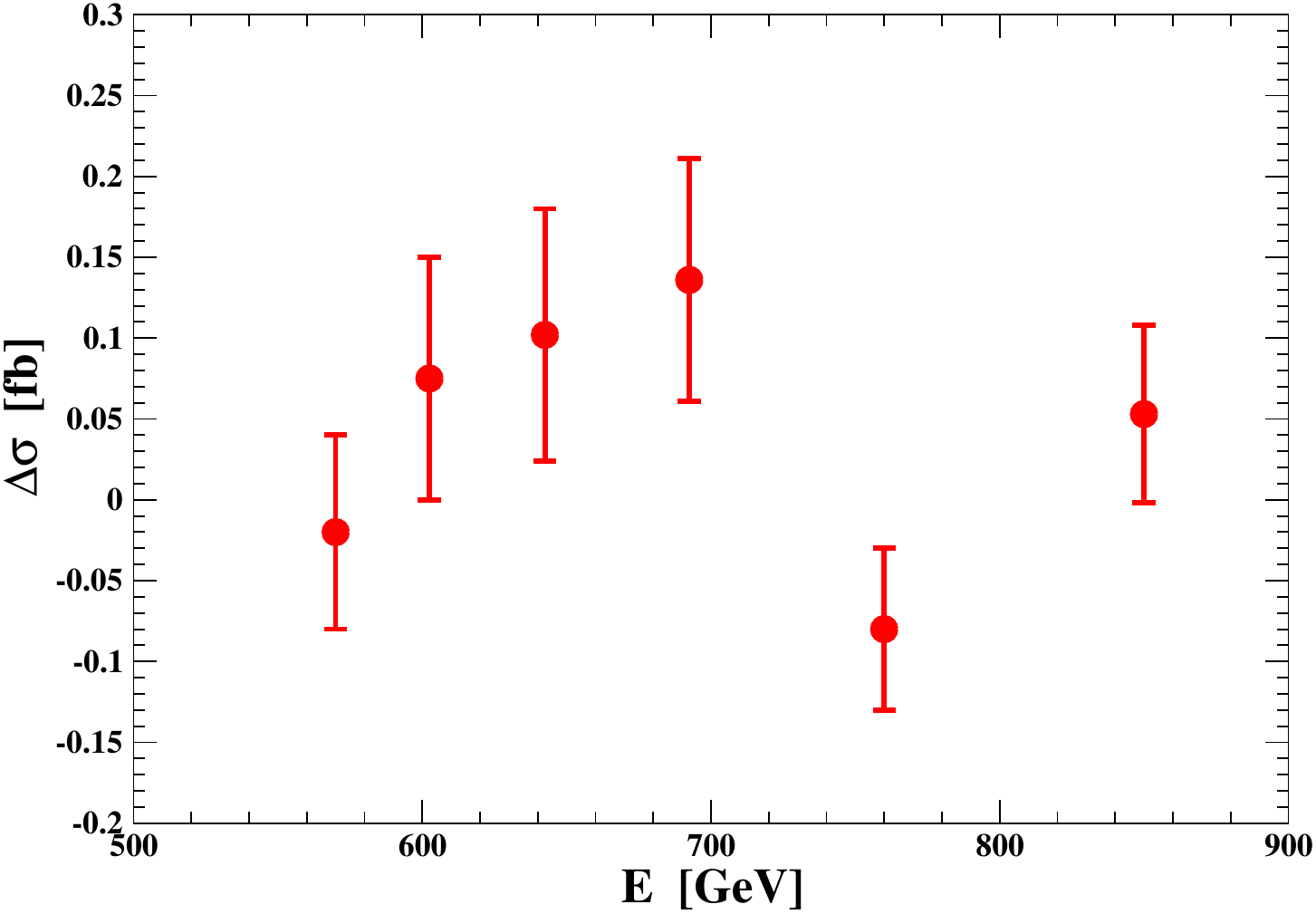}
\caption{The quantity $\Delta\sigma=(\sigma_{\rm EXP}-{\sigma}_{\rm B})$
reported for each bin in the last column of Table~\ref{leptonxsection},
for the ATLAS data of Ref.~\cite{atlasnew}.}
\label{Deltasigma}
\end{figure}
\begin{table*}
\caption{The observed ATLAS \cite{atlasnew} cross section and the estimated
background in the range of four-lepton invariant mass $m(4l)\equiv E$ from
555 to 900~GeV. These values have been obtained by multiplying the bin size
with the average differential cross sections $\langle (d \sigma/d E)\rangle$,
reported for each bin in the companion HEPData file. Besides the
non-resonant $gg\to 4l$ process, the background cross section $\sigma_B$
contains the dominating contributions from $ q\bar q \to 4l$ events
(as well as from other sources).}.
\begin{center}
\begin{tabular}{cccc}
Bin [GeV]& ${\sigma}_{\rm EXP} $~[fb]  & $\rm{ \sigma}_{\rm B} $~[fb] &
$(\sigma_{\rm EXP} - \rm {\sigma}_{\rm B} $)~[fb]  \\
\hline \hline
555--585 & 0.252 $\pm 0.056$ & $0.272 \pm 0.023$ & $-0.020 \pm 0.060$  \\
\hline
585--620 & $0.344\pm 0.070$ & $ 0.259 \pm 0.021 $ & $+0.085 \pm 0.075$ \\
\hline
620--665 & $0.356 \pm 0.075$ & $ 0.254 \pm 0.023$ & $+0.102\pm 0.078$  \\
\hline
665--720 & $0.350 \pm 0.073$ &$ 0.214 \pm 0.019$  &$+0.136 \pm 0.075$ \\
\hline
720--800 & $0.126 \pm 0.047$ &$ 0.206 \pm 0.018$ & $ -0.080 \pm 0.050$ \\
\hline
800--900 & $0.205\pm 0.052$ &$ 0.152 \pm 0.017$  & $+0.053 \pm 0.055$  \\
\hline
\end{tabular}
\end{center}
\label{leptonxsection}
\end{table*}

After having recalled this first comparison from Refs.\ \cite{CFF,universe},
we will now illustrate the indications obtained from the other ATLAS paper
\cite{atlasnew}, in which the differential four-lepton cross section
$\langle d \sigma/d E\rangle$, with $E=m(4l)$, is reported in the same energy
region. By inspection of Fig.~5 of this Ref.~\cite{atlasnew}, one finds the
same type of excess-defect sequence as in Table~\ref{leptontable}
and so additional support for the idea of a new resonance. To make
this clear, the corresponding data are given in Table~\ref{leptonxsection}
and displayed in Fig.~\ref{Deltasigma}.

Most notably, however, by comparing with Ref.~\cite{atlasnew} we can
also sharpen our analysis. The point is that the background
estimated by ATLAS, for the class of ggF-low events considered
above, contains more events than those which in principle can
interfere with our resonance. In particular, it contains a large
contribution from $ q\bar q \to 4l$ processes. Although the initial
state is $pp$ in all cases, our $H$ resonance would mainly be
produced through gluon-gluon fusion and therefore, strictly
speaking, the interference should only be computed with the
non-resonant $gg \to 4l$ background.

Obtaining this refinement is now possible because, in the HEPData
file of Ref.~\cite{atlasnew}, the individual contributions to the
expected background are reported separately. Denoting by
$\rm{\sigma}^{\rm gg}_{\rm B} $ the pure non-resonant $gg\to 4l$
background cross section, we can thus consider a corresponding
experimental cross section ${\hat \sigma}_{\rm EXP}$ after
subtracting preliminarily the ``non-ggF'' background, i.e.,
\BE
{\hat \sigma}_{\rm EXP} \; = \; \sigma_{\rm EXP} - (\rm
{\sigma}_{\rm B} - \rm{\sigma}^{\rm gg}_{\rm B} ) \; .
\label{sigmahat}
\EE
The corresponding values for these redefined cross sections and
background are given in Table~\ref{redefinedxsection}.

\begin{table*}
\caption{The ATLAS \cite{atlasnew} experimental cross section
${\hat \sigma}_{\rm EXP}$ from Eq.~(\ref{sigmahat}) for each energy bin.
The two cross sections ${\sigma}_{\rm EXP} $ and $\rm{ \sigma}_{\rm B} $
are given in Table~\ref{leptonxsection}. The other background
cross section $\rm{\sigma}^{\rm gg}_{\rm B} $ only takes into
account the non-resonant $gg\to 4l$ process and was computed by
multiplying the bin size with the average differential cross
section $(d \rm{\sigma}^{\rm gg}_{\rm B}/dE)$ in each bin. The
central value of $ {\hat \sigma}_{\rm EXP} $ in the second column of
the $720$--$800$~GeV bin is negative, because the expected background in
Table~\ref{leptonxsection}, from $q\bar q \to 4l$ events (as well as
from other sources), is larger than the experimental value itself.}
\begin{center}
\begin{tabular}{cccc}
Bin [GeV] & ${\hat \sigma}_{\rm EXP} $~[fb] &
$\rm{\sigma}^{\rm gg}_{\rm B} $~[fb]  &
$(\hat\sigma_{\rm EXP} - \rm{\sigma}^{\rm gg}_{\rm B})$~[fb]  \\
\hline \hline
555--585& 0.003 $\pm 0.060$ & $0.023 \pm 0.004$  & $ -0.020 \pm 0.060 $  \\
\hline
585--620 & $0.105 \pm 0.073$ &$ 0.020 \pm 0.003$ & $+0.085  \pm 0.075$  \\
\hline
620--665&  $0.121 \pm 0.078$ &$ 0.019 \pm 0.003$ & $+0.102  \pm 0.078$  \\
\hline
665--720 & $0.152\pm 0.075$ &$ 0.016 \pm 0.003$ & $+0.136 \pm 0.075$  \\
\hline
720--800& $-0.067\pm 0.050$ &$ 0.013 \pm 0.002$ & $-0.080  \pm 0.050$  \\
\hline
800--900 & $0.062\pm 0.055$ &$ 0.009 \pm 0.002$ & $+0.053  \pm 0.055$  \\
\hline
\end{tabular}
\end{center}
\label{redefinedxsection}
\end{table*}

\begin{table*}
\caption{Comparing the ATLAS \cite{atlasnew} cross section of
Table~\ref{redefinedxsection} with the theoretical Eq.~(\ref{sigmat}) for
the optimal set of parameters $M_H= 677$~GeV, $\Gamma_H= 21$~GeV,
$\sigma_R=0.40$~fb.}.
\begin{center}
\begin{tabular}{cccc}
Bin [GeV] & ${\hat \sigma}_{\rm EXP} $~[fb] & $ \sigma_T $~[fb] & $\chi^2$
\\ \hline \hline
555--585& 0.003 $\pm 0.060$ & 0.048 & 0.56  \\
\hline
585--620& $0.105 \pm 0.073$ & 0.056 &  0.45  \\
\hline
620--665& $0.121 \pm 0.078$ & 0.123 &   0.00  \\
\hline
665--720& $0.152\pm 0.075$ & 0.152 &   0.00  \\
\hline
720--800& $-0.067\pm 0.050$ & 0.002 &   1.90  \\
\hline
800--900 & $0.062\pm 0.055$ & 0.004 &   1.11 \\
\hline
\end{tabular}
\end{center}
\label{comparexsection}
\end{table*}

We then compare the resulting experimental ${\hat \sigma}_{\rm EXP}$ with
the theoretical $\sigma_T$ from Eq.~(\ref{sigmat}), after the
identification $\sigma_b= \rm{\sigma}^{\rm gg}_{\rm B}$. By
parametrising the ATLAS differential background
$(d \rm{\sigma}^{\rm gg}_{\rm B}/dE) \sim A\times({\rm 710~GeV}/E)^{\nu}$
with $A\sim (2.42 \pm 0.18)\times 10^{-4}$ fb/GeV  and $\nu\sim 5.24
\pm 0.45$, and integrating the various contributions to
Eq.~(\ref{sigmat}) within each energy bin, a fit to the data results in
$M_H= 677^{+30}_{-14}$~GeV, $\Gamma_H= 21^{+28}_{-16}$~GeV, and
$\sigma_R= 0.40^{+0.62}_{-0.34}$ fb. The comparison for the optimal
parameters is shown in Table~\ref{comparexsection}.

As in the case of the ggF-low events, the quality of our fit is good,
but error bars are large. Still, by restricting ourselves again to the
central values, we find a good agreement with our expectations. Indeed, by
rescaling the partial width given in Eq.~(\ref{rel1}) (Sec.~3), from
$\Gamma(H\to ZZ)\sim 1.6$~GeV down to 1.55~GeV (for a mass $M_H$ from 700
to 677~GeV), and fixing $\Gamma_H$ at its central value of 21~GeV, we find a
branching ratio $B(H\to ZZ)\sim 0.073$. For the central value
$\sigma^{\rm ggF} (pp\to H)\sim 1100$~fb from Ref.~\cite{yellow} at
$M_H = 677$~GeV, this would then imply a theoretical peak cross
section $(\sigma_R)^{\rm Theor}=1100\times0.073\times0.0045\sim0.36$~fb,
which only differs by 10\%  from the central value
$\langle\sigma_R\rangle= 0.40$~fb of our fit. Also, from the central
values of the fit $\langle\sigma_R\rangle= 0.40$~fb and
$\langle\gamma_H\rangle= 0.031$, we find
$\langle\sigma_R\rangle \times \langle\gamma_H\rangle\sim 0.012$~fb,
again in good agreement with our Eq.~(\ref{exp34}).

Let us now summarise these results. By considering the two ATLAS papers
\cite{atlas4lHEPData} and \cite{atlasnew}, we have found consistent
indications of a new resonance in our theoretical mass range
$(M_H)^{\rm Theor}\sim 690\,(30)$~GeV. In particular, by comparing
with the cross-section data of Ref.~\cite{atlasnew}, we have
identified more precisely the non-resonant background $gg\to 4l$,
which can interfere with a second resonance $H$ produced mainly via
gluon-gluon fusion. In this sense, the determinations obtained with
our Eq.~(\ref{sigmat}) are now more accurate, from a theoretical
point of view. In practice, there is not much difference with the
previous analysis \cite{CFF,universe} based on the ggF-low events of
Ref.~\cite{atlas4lHEPData}. Indeed, the two mass values
$(M_H)^{\rm EXP}=677^{+30}_{-14}$~GeV vs.\ $(M_H)^{\rm EXP}= 706\,(25)$~GeV
\cite{CFF,universe} and decay widths $\Gamma_H= 21^{+28}_{-16}$ GeV vs.\
$\Gamma_H=29\pm 20$~GeV \cite{CFF,universe}, are compatible within their
rather large experimental uncertainties. Most notably, our crucial
correlation in Eq.~(\ref{exp34}) is well reproduced by the central values
of the fits to the two date sets.

\subsection{The ATLAS high-mass $\gamma\gamma$ events}

\begin{figure}[htb]
\includegraphics[trim = 0mm 0mm 0mm 0mm,clip,width=8.6cm]
{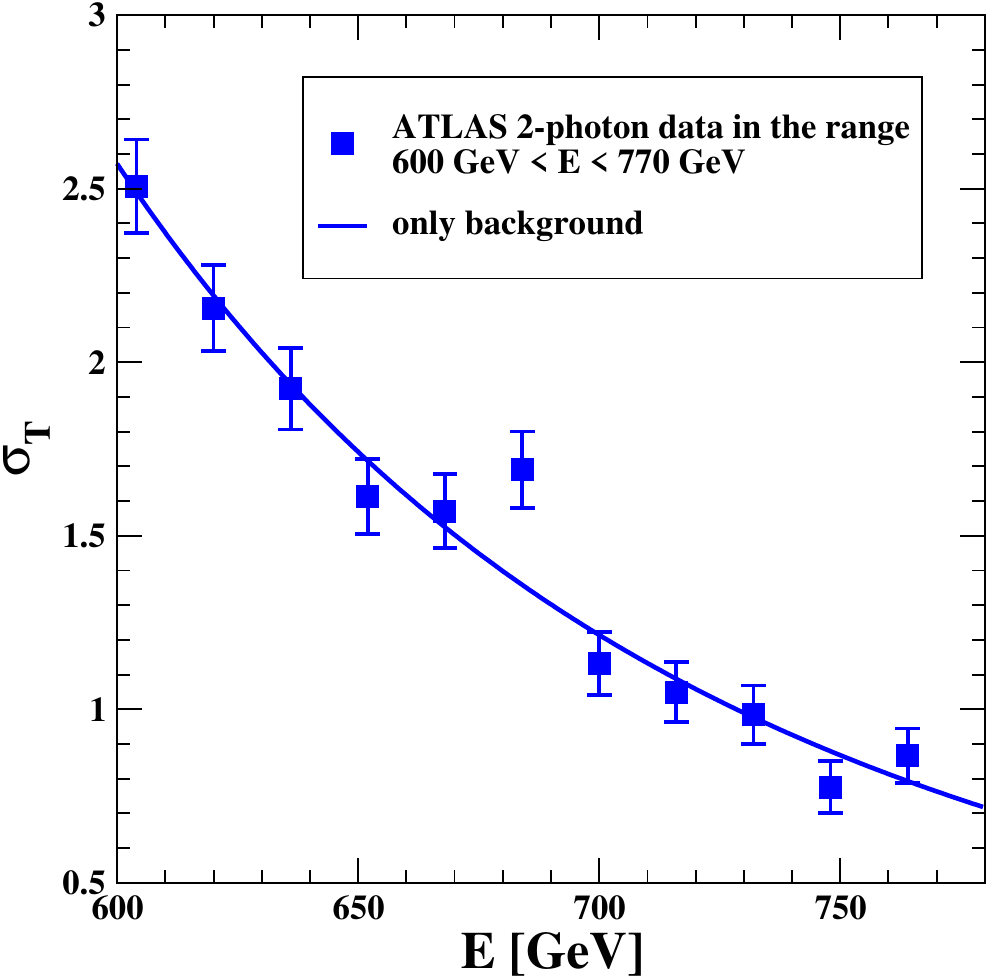}
\caption{The fit with Eq.~(\ref{sigmat}) and $\sigma_R=0$ to the ATLAS
\cite{atlas2gammaplb} data in Table~\ref{2gammatable}, converted to cross
sections in fb. The chi-squared value is $\chi^2=14$, with the background
parameters $A=1.35$~fb and $\nu=4.87$.}
\label{twogamma0}
\end{figure}

\begin{table*}
\caption{The ATLAS \cite{atlas2gammaplb} number of events $N=N(\gamma\gamma)$,
in bins of 16~GeV and for a luminosity of 139~fb$^{-1}$, in the range of
invariant mass $\mu=\mu(\gamma\gamma)=600\div 770$~GeV. These values
were extracted from Fig.~3 of Ref.~\cite{atlas2gammaplb}, because the
corresponding numbers are not reported in the companion HEPData file.
In the fits, statistical errors were assumed to be given by $\sqrt{N}$,
as for a Poisson distribution.}
\begin{center}
\begin{tabular}{cccccccccccc}
$\mu$ & 604 & 620 & 636 & 652 & 668 & 684 & 700 & 716 & 732 & 748 & 764 \\
\hline $N$ & 349 & 300 & 267 & 224 & 218 & 235 & 157 & 146 & 137 & 108 & 120\\
\hline
\end{tabular}
\end{center}
\label{2gammatable}
\end{table*}

\begin{figure}[htb]
\includegraphics[trim = 0mm 0mm 0mm 0mm,clip,width=8.6cm]
{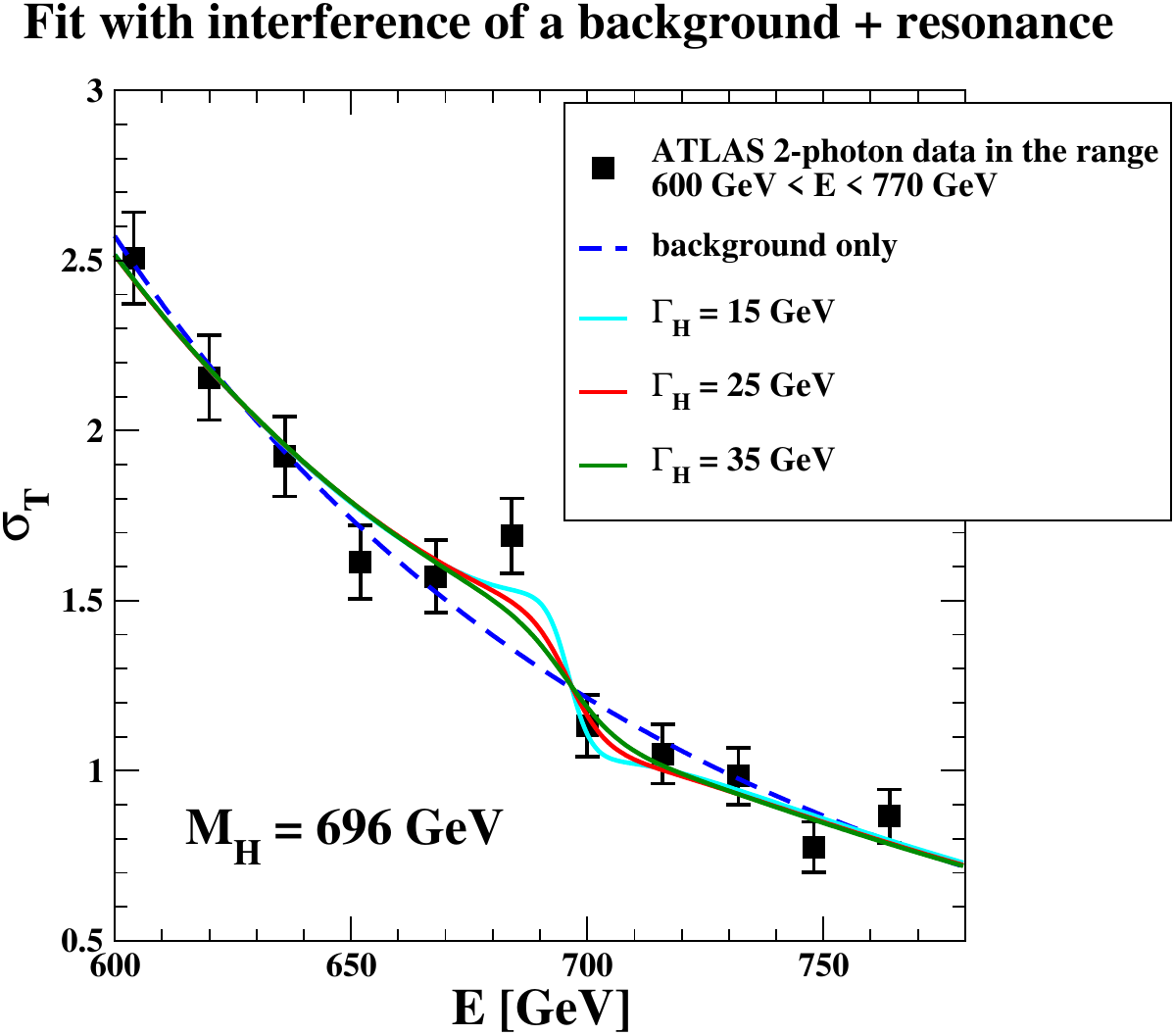}
\caption{Three fits with Eq.~(\ref{sigmat}) to the ATLAS \cite{atlas2gammaplb}
data in Table~\ref{2gammatable}, converted to cross sections in fb.
The $\chi^2$ values are 7.5, 8.8, and 10.2, for $\Gamma_H= 15$, 25,
and 35~GeV, respectively.}
\label{twogamma1}
\end{figure}

Searching for other signals, in Refs.~\cite{CFF,universe} one
considered the distribution of the inclusive diphoton production by
ATLAS \cite{atlas2gammaplb} in the range of invariant mass $600\div770$~GeV.
The corresponding entries in Table~\ref{2gammatable} were extracted from
Fig.~3 of Ref.~\cite{atlas2gammaplb}, because the numerical values are
not reported in the companion HEPData file. By parametrising the
background with a power-law form
$\sigma_B(E)\sim A\times 685 \, ({\rm GeV}/E)^{\nu}$, doing a fit to the
data in Table~\ref{2gammatable} gives a good description of all data
points with the exception of a sizeable excess at 684~GeV (estimated by
ATLAS to have a local significance of more than $3\sigma$); see
Fig.~\ref{twogamma0}. This illustrates how a relatively narrow resonance
might remain hidden behind a large background almost everywhere, the main
signal being just a small interference effect. For this reason, with the
exception of the mass $M_H=696\,(12)$ GeV, the resonance parameters are
determined only very poorly. As for the total width, one finds
$\Gamma_H= 15^{+18}_{-13}$~GeV, which is consistent with the other loose
determination $\Gamma_H = 21^{+28}_{-16}$~GeV from the four-lepton data.
In Fig.~\ref{twogamma1} we show three fits with Eq.~(\ref{sigmat}), viz.\
for $\Gamma_H=15$, 25, and 35~GeV. The widths vary substantially, but the
curves cross the background at the same point $M_H=696$~GeV where the
interference vanishes. Concerning the peak cross section
$\sigma_R=\sigma_R(pp\to H\to \gamma\gamma)$, the fit produces
$\sigma_R=0.025^{+0.055}_{-0.023}$~fb, with central value
$\sqrt{\sigma_R \, \sigma_B(684)}\sim 0.18$~fb or about $+26$
events. To have an idea, for $M_H\sim 700$~GeV, $\Gamma_H\sim 29$~GeV
and $\sigma(pp\to H)\sim 1$ pb, the partial width
$\Gamma(H \to \gamma\gamma)\sim 29$~keV of Ref.~\cite{widths} implies
$B(H\to \gamma\gamma)\sim 1\times 10^{-6}$ and a peak cross section
$\sigma_R(pp\to H\to \gamma\gamma)\sim 0.001$~fb. Instead, with a
larger two-photon width (see footnote~12) one could start to
approach the lower band of the fit.

In conclusion, the localised $3\sigma$ excess at 684~GeV admits two
different interpretations:
\begin{enumerate}[a)]
\item
a statistical fluctuation above a pure background, see
Fig.~\ref{twogamma0};
\item
the signal of a heavy, relatively narrow resonance, see
Fig.~\ref{twogamma1}.
\end{enumerate}

\subsection{ATLAS and CMS $ (b \bar b + \gamma\gamma)$ events}
\begin{figure}[htb]
\includegraphics[trim = 0mm 0mm 0mm 0mm,clip,width=8.6cm]
{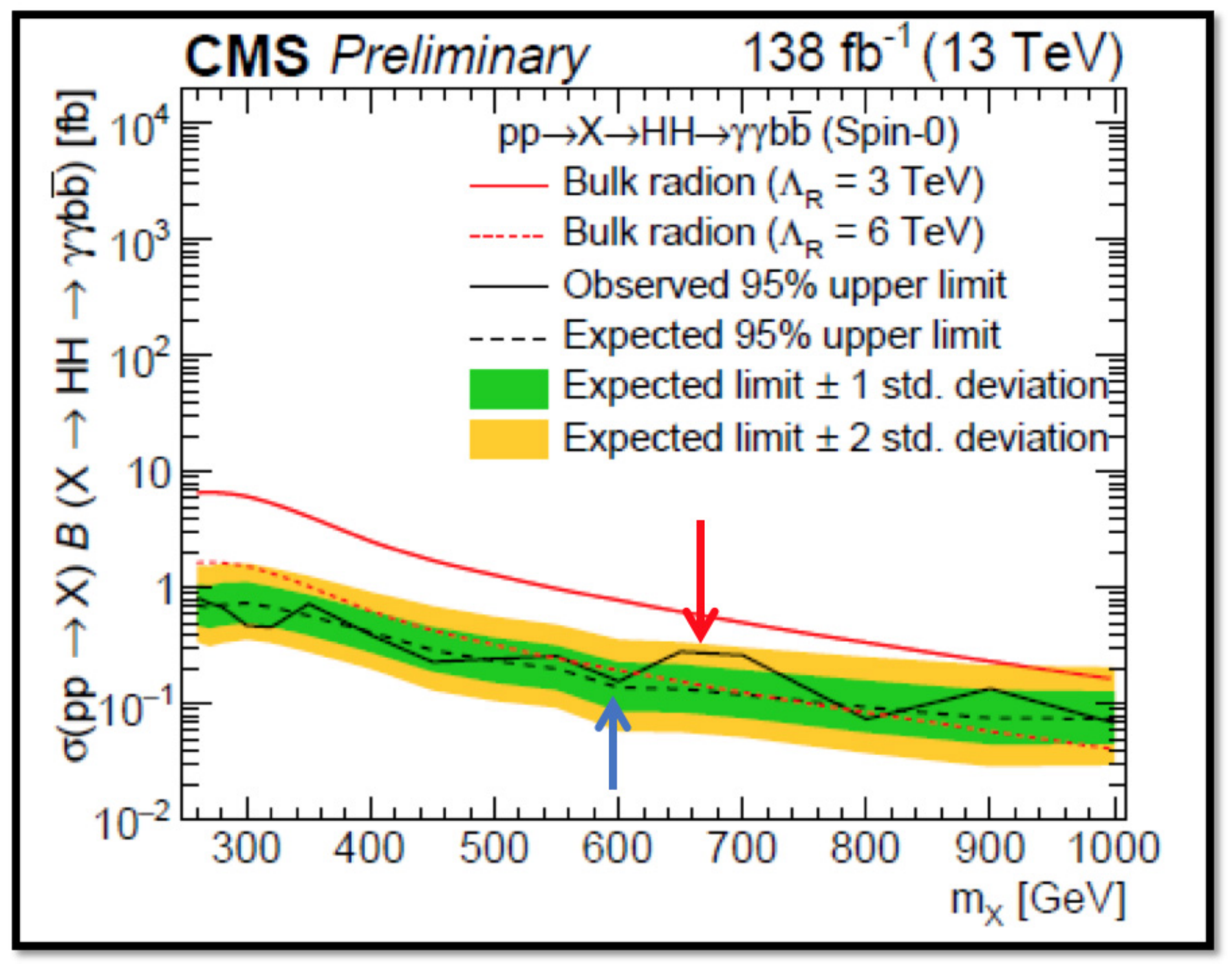}
\caption{Expected and observed 95\% upper limit for the cross
section $\sigma (pp\to X\to h(125)h(125)\to b\bar b+\gamma\gamma)$
observed by the CMS Collaboration \cite{CMS-PAS-HIG-21-011}.}
\label{CMS_DBGG}
\end{figure}
The ATLAS and CMS Collaborations have also searched for
new resonances decaying, through a pair of $h=h(125)$ scalars, into
the peculiar final state made up of a $b\bar b$ quark pair and a
$\gamma\gamma$ pair. In particular, in Ref.~\cite{CMS-PAS-HIG-21-011}
the cross section for the full process
\BE
\sigma({\rm full}) \; = \; \sigma(pp\to X\to hh\to b\bar b+\gamma\gamma)
\EE
was considered. For a spin-zero resonance, the 95$\%$ upper limit
$\sigma({\rm full})<0.16$~fb, for an invariant mass of 600~GeV, was found
to increase by about a factor of two, up to $ \sigma({\rm full})<0.30$~fb
on a plateau  $650\div 700$ GeV and then to decrease at larger
energies; see Fig.~\ref{CMS_DBGG}.
\begin{figure}[htb]
\includegraphics[trim = 0mm 0mm 0mm 0mm,clip,width=8.6cm]
{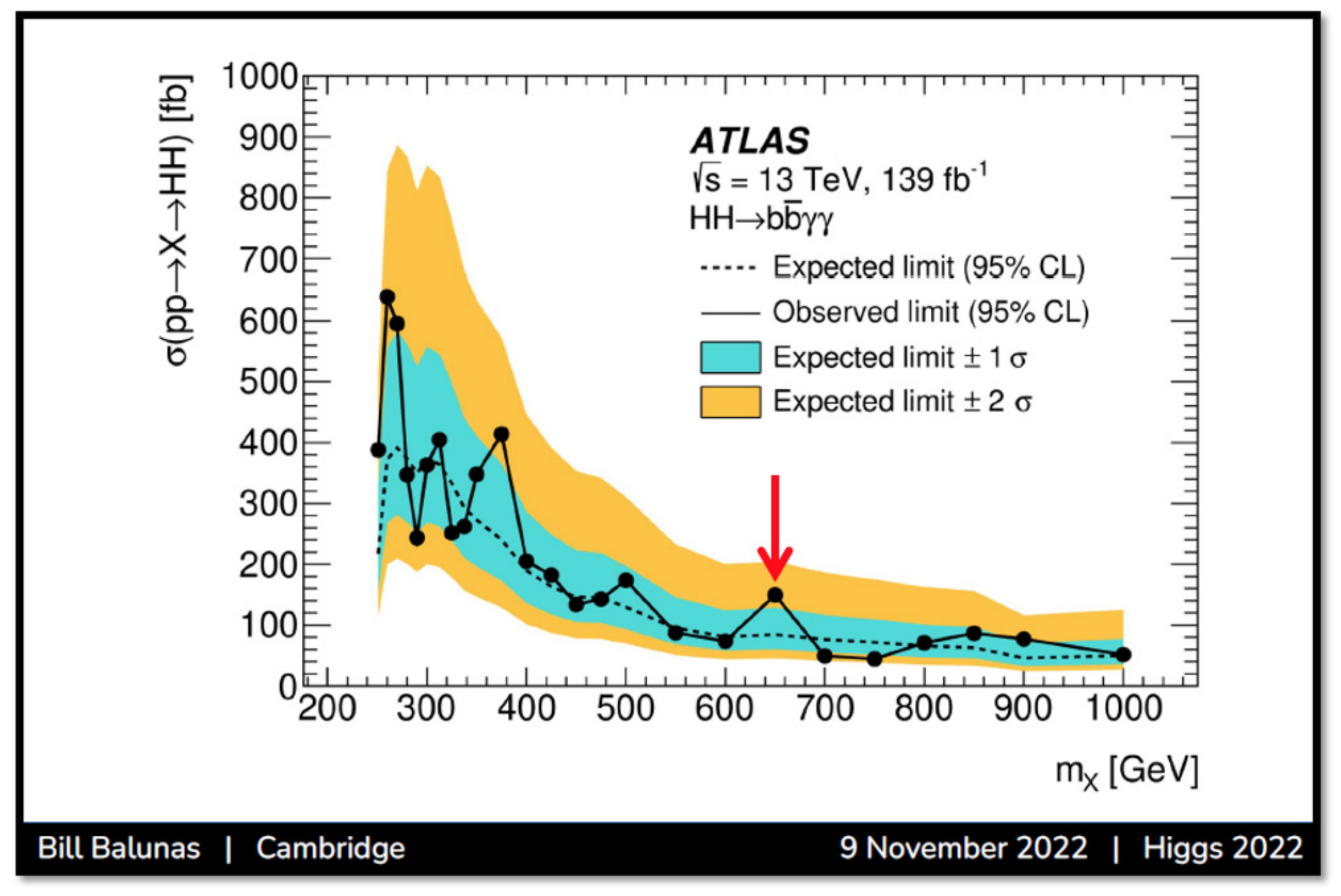}
\caption{Expected and observed 95$\%$ upper limit for the cross
section $\sigma (pp\to X \to h(125)h(125))$ extracted by ATLAS
\cite{ATLAS_BBGG_paper} from the final state $(b \bar b + \gamma\gamma)$.
The figure is taken from the talk given by Bill Balunas at ``Higgs 2022''
and is the same as Fig.~15 in Ref.~\cite{ATLAS_BBGG_paper}.}
\label{ATLAS_BBGG}
\end{figure}

The local statistical significance of the excess is modest, about
$+1.5\sigma$, but the relevant mass region is precise and agrees
well with our prediction. Therefore, we could compare this observed
excess around 650~GeV, which from Fig.~9 is about 0.15 fb, with the order
of magnitude of the peak cross section expected for the $H$ resonance, namely
\BE
\Delta\sigma^{({\rm full})}
\sim \sigma(pp\to H)B(H\to hh)\,2B(h\!\to\!b\bar b)\,B(h\!\to\!\gamma\gamma).
\EE
From this relation and the PDG value
$2 B(h\!\to\!b\bar b)\,B(h\!\to\!\gamma \gamma) \sim 0.0026\,(4)$, the CMS
upper bound can then be re-written as 
\BE
\sigma^{\rm CMS}(pp\to H \to hh) \; < \;  (59 \pm 9)\;{\rm fb} \; .
\EE
By comparing with the corresponding expectation for the $H$ resonance,
assuming the theoretical branching ratio $B^{\rm TH}(H\to hh) \sim 0.046$
of Sec.~3 and our ggF reference value $\sigma(pp\!\to\!H)=$ 1090\,(170) fb,
we find 
\BE
\label{THbbgg}
\sigma^{\rm TH}(pp\to H \to hh) \; \sim \; (50 \pm 8) \; {\rm fb} \; ,
\EE
consistently with the above CMS upper limit. 

The analogous ATLAS plot is shown in Fig.~\ref{ATLAS_BBGG} (being
the same as Fig.~15 of the ATLAS paper in Ref.~\cite{ATLAS_BBGG_paper}).
Again, one finds a modest $+1.2\sigma$ excess at $650\,(25)$~GeV, followed
by two \/consecutive $-1.4\sigma$ defects, viz.\ at 700\,(25) and 750\,(25)
GeV, suggestive of a negative above-peak $(M^2_H - s)$ interference effect,
as found in the ATLAS four-lepton data. The mass would thus lie between 650
and 700~GeV, where the interference effect changes sign. From the observed
upper limit of 150~fb, with an estimated background of 82~fb, we find an
excess $\sigma^{\rm ATLAS}(pp\to H\to hh)<68$~fb, in very good agreement
with our theoretical value in Eq.~(\ref{THbbgg}). 

Since the three-body decays $H\to hhh$, $H\to hW^+W^-$, and $H\to hZZ$ should
only give a modest contribution to the total width, from the estimates in
Sec.~3 and the consistency with the theoretical value
$B^{\rm TH}(H\to hh) \sim 0.046$, we deduce an upper limit for the total
width $\Gamma_H < 35$ GeV.

\subsection{CMS $\gamma\gamma$ events produced in $pp$
diffractive scattering}

\begin{figure}[htb]
\includegraphics[trim = 0mm 0mm 0mm 0mm,clip,width=8.6cm]
{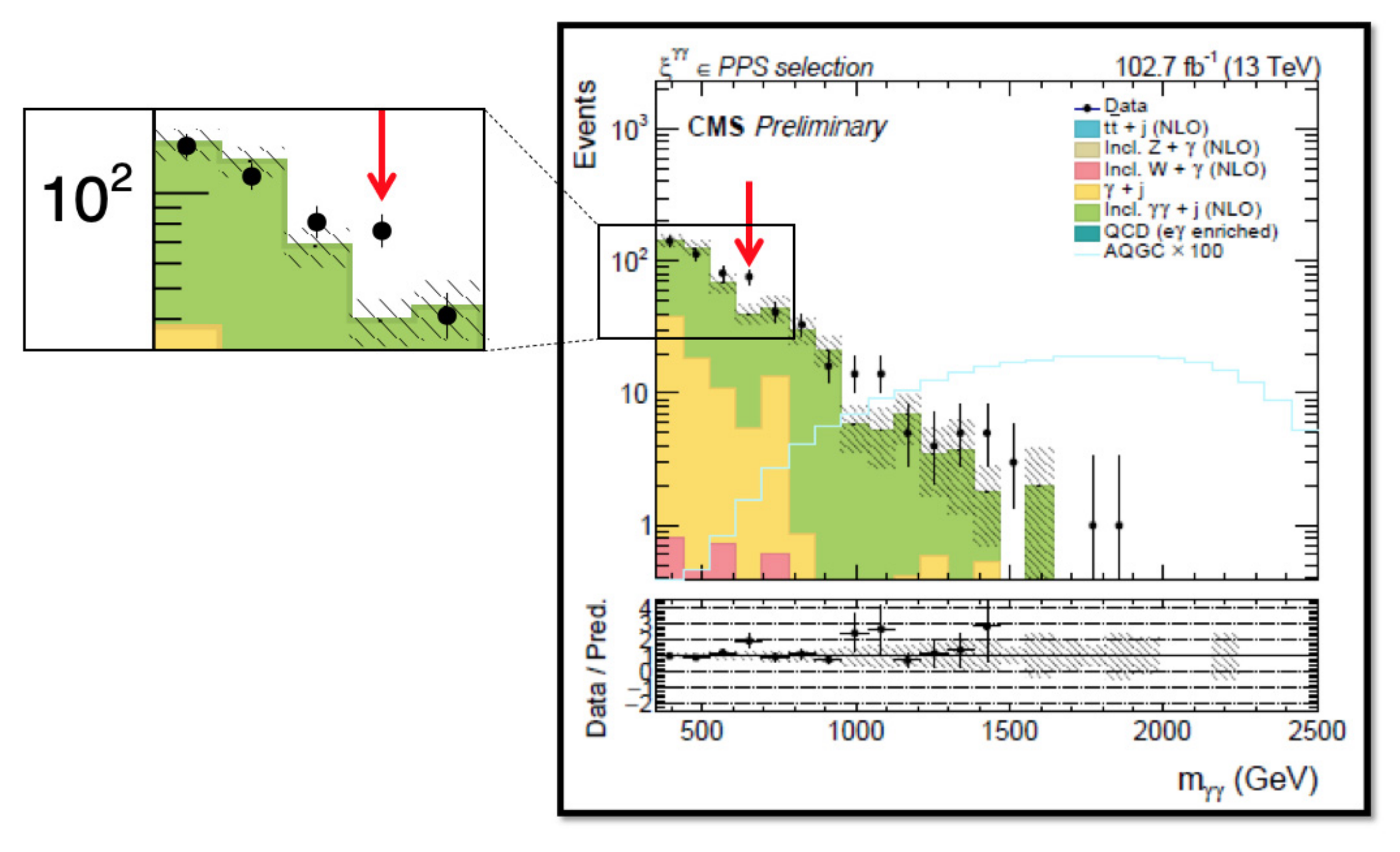}
\caption{The number of $\gamma\gamma$ events
produced in $pp$ diffractive scattering, as reported by CMS
\cite{CMS-PAS-EXO-21-007}. In the range $650\,(40)$~GeV, the
observed number was $N_{\rm obs}\sim 76\,(9)$, to be compared to an
estimated background $N_{\rm B}\sim 40\,(9)$.}
\label{diffractive}
\end{figure}

Finally, the CMS and TOTEM Collaborations have been searching for
high-mass photon pairs produced in $pp$
diffractive scattering, i.e., when both final protons are
tagged and have large $x_F$. For our purpose, the relevant
information is contained in Fig.~\ref{diffractive}, taken from
the CMS report Ref.~\cite{CMS-PAS-EXO-21-007}. In the range of
invariant mass $650\,(40)$~GeV and for a statistics of 102.7~fb$^{-1}$,
the observed number of $\gamma\gamma$ events is $N_{\rm obs}\sim76\,(9)$,
to be compared with an estimated background $N_{\rm B}\sim 40\,(9)$,
which is quoted as being the best estimate by CMS. In the most
conservative case, namely $N_{\rm B}=49$, this is a local $3\sigma$
effect and the only statistically significant excess in the
plot.\footnote{We emphasise that a lower-statistics
paper was published by the CMS and TOTEM Collaborations in
Ref.~\cite{CMS_TOTEM}. The latter analysis was based on the
statistics collected in 2016, with an integrated luminosity of
9.4~fb$^{-1}$. Instead, Fig.~5.d of Ref.~\cite{CMS-PAS-EXO-21-007},
reported here as our Fig.~\ref{diffractive}, is based on the full
RUN2 data collected in 2016, 2017, and 2018, with an integrated
luminosity of 102.7 fb $^{-1}$ and so more than ten times larger.}

\subsection{Experimental summary}

Let us finally summarise our review of LHC data:
\begin{enumerate}[i)]
\item
The ATLAS ggF-like four-lepton events in Table~\ref{leptontable}
show deviations from the background with a definite excess-defect
sequence, which could indicate the presence of a resonance. The same
pattern is also visible in the ATLAS data for the corresponding
differential cross section; see Fig.~5 of Ref.~\cite{atlasnew} and
the corresponding integrated cross sections in Tables~\ref{leptonxsection}
and \ref{redefinedxsection}. From the last
column of these Tables, the combined statistical significance of the
observed deviations can be estimated at the $3\sigma$ level. A fit with
Eq.~(\ref{sigmat}) gives a good description of the data for a
resonance mass $M_H= 677^{+30}_{-14}$ GeV; see
Table~\ref{comparexsection}.
\item
Observing the $+3\sigma$ excess around $684$~GeV  in the
inclusive ATLAS $\gamma\gamma$ events, a fit to these data was
performed in Refs.~\cite{CFF,universe}. The resulting mass was
$M_H=696\,(12)$~GeV. \\
Furthermore, one should note:
\item
Measurements of the ($b\bar b + \gamma\gamma)$ final state, by both ATLAS
and CMS, indicate various types of deviations from the background in the region
$650\div750$ GeV. From the observed excesses/defects of events, the
mass of the hypothetical new resonance could be roughly estimated to be
$M_H = 675\,(25)$ GeV. The overall, local statistical significance is about
2 $\sigma$.
\item
The CMS-TOTEM $\gamma\gamma$ events produced in $pp$ diffractive scattering
indicate a $+3\sigma$ excess in the region of invariant mass
$M_H=650\,(40)$~GeV.
\end{enumerate}

Since the above determinations i)-iv) are well aligned within the
uncertainties, we can try to combine the mass values and obtain
$(M_H)^{\rm comb}\sim 685\,(10)$~GeV, in very good agreement with our
prediction $(M_H)^{\rm Theor}=690(30)$~GeV.
We emphasise again that, when comparing with a definite
prediction, one should look for deviations from the background
nearby, so that local significance is not downgraded by the so called
``look elsewhere'' effect. Therefore, since the correlation of the above
measurements is small, one could also argue that the combined
statistical evidence for a new resonance in the expected mass range
is close to (if not above) the traditional $5\sigma$ level. Certainly,
one can conclude that the present situation is unstable and could soon
be resolved by including new LHC data. In this way, the present non-negligible
statistical significance could become an important discovery. In view of these
possible future developments, we will illustrate in the following Sec.~5 the
basic ingredients of a coupled-channel calculation that could be useful to
further refine the theoretical predictions for the mass and width of the
hypothetical new resonance, when interacting with the gauge and
fermion sectors of the Standard Model.

\section{Unitarity constraints from coupled-channel calculations}
\label{rse-intro}

Our estimate $(M_H)^{\rm Theor} \sim$ 690(30)
GeV was obtained by combining analytical and numerical results in a
pure $\Phi^4$ theory. In this sense, this mass value should be
considered a ``bare'' mass that would become the physical mass, to
be compared to experiment, after including the interactions with the
gauge and fermion fields. For this reason, it is remarkable that the
phenomenological analysis illustrated in Sec.~4 suggests a new
resonance precisely in the expected region of this bare mass.

As for the total decay width, we started from the standard
calculations for a single Higgs particle of mass $M_H \sim$ 700~GeV,
by taking into account the strong renormalisation of the
conventional large widths into longitudinal $W$s by a factor
$(125/700)^2\sim $ 0.032. This suppression leads to the estimates in
Eqs.~(\ref{rel1},\ref{rel2}) and to the precise correlation in the
four-lepton channel in Eq.~(\ref{exp34}), which, within the large
uncertainties of the fit to the ATLAS four-lepton data, is reproduced
by the central values $\langle\sigma_R(pp\to H \to 4l)\rangle \sim
0.40$~fb and $\langle\gamma_H\rangle\sim 0.031$. If checked to high
accuracy with future data, this correlation would imply that,
indeed, the new resonance is the second resonance of the Higgs
field.

Still Eq.~(\ref{exp34}) is, to a large extent, independent of the
total width. Our expectation $\Gamma^{\rm Theor}(H\to all)= 31\div
35$~GeV is consistent with the very loose determinations obtained
from the fit to the ATLAS four-lepton and $\gamma\gamma$ events.
However, with more precise measurements, things could change
significantly. To this end, let us consider the following
hypothetical experimental values: $(M_H)^{\rm exp}=700\,(10)$~GeV,
$\Gamma^{\rm exp}(H\to all)=22\,(3)$~GeV, and
$\sigma_R=[\sigma_R(pp\to H \to 4l)]^{\rm exp}=0.33\,(5)$~fb. These
values would imply a correlation $[\gamma_H\times \sigma_R]^{\rm
exp}=0.010\pm 0.002$, in excellent agreement with Eq.~(\ref{exp34}),
and thus with Eqs.~(\ref{rel1},\ref{rel2}). But the total width
would lie at three-sigma from the expected theoretical value
$\Gamma^{\rm Theor}(H\to all) \sim 33$~GeV of Sec.~3. Then, the most natural explanation would consist
in a substantial defect in the main partial width $\Gamma(H \to t
\bar t)$. Here, for $m_t \sim 173$~GeV one has the lowest-order
estimate
\begin{equation}
\label{perturbed} \Gamma^{\rm LO} (H\to t \bar t)  =
\frac{3 M_H}{8\pi} \, \frac{m^2_t}{v^2} \,
\left[1-4\frac{m^2_t}{M^2_H}\right]^{3/2} \sim \; 27.1 \, {\rm GeV} \; ,
\end{equation}
very close to the full value $\Gamma(H\!\to\!t\bar t)\!\sim\!27.5$~GeV
in Ref.~\cite{handbook}.

Differently from the more common situation where a decay width turns
out to be larger than expected, which suggests the presence of new
final states, the presence of a substantial defect could indicate
the suppression produced by the unitarity constraints in the
presence of several competing decay channels. This problem can be
addressed in a coupled-channel analysis such as the so-called
Resonance-Spectrum-Expansion (RSE) model. It was developed in
Ref.~\cite{hepph0304015,AOP324p1620,PRD80p094011} as a
momentum-space version of a much earlier
\cite{PRD27p1527,ZPC30p615} coupled-channel model for
unquenched meson spectroscopy formulated in coordinate space, which
accounts for surprisingly large dynamical effects of strong decay on
the spectra.

In Subsec.~\ref{rse-model} the RSE model is briefly revisited and
in Subsec.~\ref{rse-Higgs} its possible application to a
heavy-Higgs system is sketched.

\subsection{RSE model for non-exotic meson-meson scattering}
\label{rse-model}
\begin{figure*}[htb]
\includegraphics[trim = 39mm 214mm 20mm 44mm,clip,width=17.2cm,angle=0]
{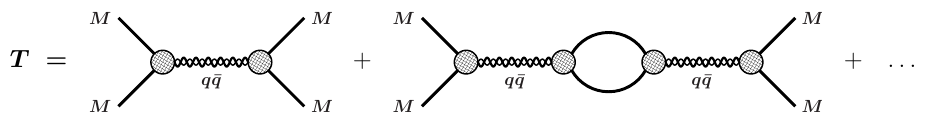}
\caption{Graphical depiction of RSE $T$-matrix for non-exotic two-meson
scattering.}
\label{rse-graphs}
\end{figure*}

The RSE model is a very efficient multichannel formalism that guarantees
$S$-matrix unitarity and analyticity in a non-perturbative description of
non-exotic meson resonances and bound states as quark-antiquark ($q\bar{q}$)
systems coupled to a variety of meson-meson ($MM$) decay channels with
the same quantum numbers. Transitions between the $q\bar{q}$ and $MM$ sectors
proceed via the creation or annihilation of a new $q\bar{q}$ pair with vacuum
quantum numbers and so in a $^{3\!}P_0$ state. Thus, the off-energy-shell RSE
$T$-matrix, graphically depicted \cite{AOP324p1620} in Fig.~\ref{rse-graphs},
describes an incoming pair of mesons that at a $^{3\!}P_0$ annihilation
vertex couple to an intermediate quark-antiquark state propagating in the
$s$-channel and then, via a $^{3\!}P_0$ creation vertex, give rise to an
outgoing  --- possibly different --- pair of mesons. The $T$-matrix in
Fig.~\ref{rse-graphs} clearly amounts to a Born term plus an infinite
set of bubble diagrams. As for the intermediate state, a whole tower of
bare $q\bar{q}$ levels with the same quantum numbers is taken. The
vertices are modeled by spherical Bessel functions, which are the Fourier
transforms of a spherical $\delta$ function, thus mimicking string
breaking at a certain interquark distance. Due to the separability of
the effective $MM$ interaction and the chosen vertex functions,
the corresponding $T$-matrix and also the on-energy-shell $S$-matrix can be
solved in closed form, both algebraically and analytically, where the
spherical Bessel functions act as regulators in the intermediate-state
$s$-channel MM loops.

\subsection{RSE model for \boldmath{$H \to t\bar{t}, W^+W^-, ZZ, hh$}}
\label{rse-Higgs}

In this subsection, we will consider a possible extension of the RSE
model to the case of the Higgs field. To this end, we will assume
the existence of two states with bare, real mass-squared $r^{(n)}$
where $n=1$ denotes the $h=h(125)$ resonance and $n=2$ the heavy
second resonance. Each bare state couples to other $k=1,\ldots,M$
states (in our case the fermion and gauge fields) through
three-point vertices. This is because the RSE approach can only deal
with intermediate two-particle states in the loops. Through this
coupling, the bare masses will become the physical ones, represented
by bound-state or resonance $S$-matrix poles in the complex-energy
plane.

\begin{figure*}[!htb]
\includegraphics[trim = 40mm 214mm 20mm 43mm,clip,width=17.2cm,angle=0]
{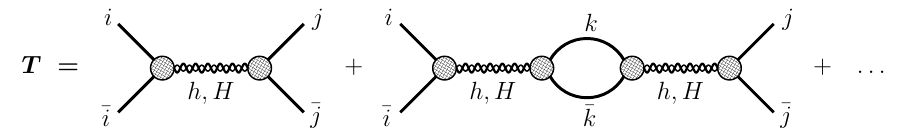}
\caption{Graphical depiction of the RSE
$T$-matrix for two Higgs-field mass states $h$ and $H$. The labels
$i,j,k$ and $\bar{i},\bar{j},\bar{k}$ at the external legs and the
loop stand for $(t/\bar{t},W^\pm,Z,h$) and $(\bar{t}/t, W^\mp,Z,h)$,
respectively.}
\label{rsehiggs-graphs}
\end{figure*}

For a heavy Higgs resonance $H$ one could consider in principle the
four main decay modes $t\bar{t}$, $W^+W^-$, $ZZ$, and $hh$, which we
could denote as $k=1, 2, 3, 4$, respectively. With such heavy
intermediate states, no imaginary part can arise for the lower
resonance $h(125)$ and its bare mass would in principle be adjusted
so that the real part of the pole has the observed, physical value
$m_h=$ 125~GeV. The RSE $T$-matrix elements could then be represented
as in Fig.~\ref{rsehiggs-graphs}.

From the relation with the S-matrix,
\begin{equation}
S_{ij}(p_i,p_j;s)\;=\;\delta_{ij}\:+\:2i \, T_{ij}(p_i,p_j;s)\;.
\label{smatrix}
\end{equation}
one gets the unitarity relations
\begin{equation}
\mathcal{I}m\,T_{ij}(p_i,p_j;s) \; = \; \sum_{n=1}^4
T_{in}^\star(p_i,p_n;s)\,T_{nj}(p_n,p_j;s)\;. \label{unitarity}
\end{equation}
It is then intuitive that the constraints placed by the simultaneous
presence of many different channels can affect the normalisation of
each matrix elements and modify the various partial decay widths. We
are aware that gauge invariance, which is usually preserved order by
order in perturbation theory, makes the application of this method
to the Standard Model problematic. Nevertheless, here we will
outline the general structure of the problem. But in
part of the actual calculations and also for comparison, we will
restrict ourselves to $t \bar t$ graphs, which form a
gauge-independent set and are expected to give the main contribution
to the $H$ width. Significant modification of $\Gamma(H\to t \bar
t)$ could still be due to the mixing with the $h$ resonance.

In Eq.(\ref{unitarity}) we consider the centre-of-mass (CM) frame, where in
terms of on-shell four-momenta $P_i\equiv (E_i,\bf {p}_i)$, $P_j\equiv
(E_j,\bf {p}_j)$, one has $s=4(p_i^2+m_i^2)=4(p_j^2+m_j^2)$. Basic
elements of the scheme are the vertices which have three indices,
say $V^{(n)}_{ii}(P_1,P_2)$. By defining $P\equiv
(\sqrt{s},0,0,0)$, expressing $P_i$, $P_{\bar i}$ as
$(\sqrt{s}/2,\pm \sqrt{s/4 - m^2_i},0,0)$, and $P_j$, $P_{\bar
j}$ as $(\sqrt{s}/2, \pm \sqrt{s/4 - m^2_j},0,0)$, we can then
introduce the effective couplings $g^{(n)}_{i}(s)\equiv \psi^*_{\bar
i}(P_{\bar i}) V^{(n)}_{i \bar i}(P_i,P_{\bar i})\psi_i(P_i) $ and
$g^{(n)}_{j}(s)\equiv \psi^*_{\bar j}(P_{\bar j})V^{(n)}_{j \bar
j}(P_j,P_{\bar j})\psi_j(P_j) $ by employing the bare RSE
propagators
\begin{equation}
\mathcal{R}^{(n)}_{ij}(s) \; = \; \frac{g^{(n)}_{i}(s)
 g^{(n)}_j (s) }{s-r^{(n)}} \; .
\label{rse}
\end{equation}
The $T$-matrix elements are then very similar to the multichannel RSE
generalisation in Ref.~\cite{PRD80p094011}:
\begin{equation}
T_{ij}(p_i,p_j;s)\; = \; \sum_{k=1}^{M}
\sum_{n=1}^{2}\mathcal{R}^{(n)}_{ik}(s)
\left\{[\One-(\Omega{R})^{(n)}(s)]^{-1}\right\}_{\!kj} \; ,
\label{tmatrix}
\end{equation}
where the loop function $\Omega_{kk}(s)$, for equal-mass
($k\bar{k}$) intermediate states can be expressed as
\begin{equation}
\label{loop}
\Omega^{nm}_{kk}(s)  =
i\!\int\!\frac{d^4q}{(2\pi)^4}\,V^{(n)}_{kk}(P,q)\,
G_k(q)\,V^{(m)}_{kk}(q,P)\,G_k(P+q)
\end{equation}
and
\begin{equation}
 \label{product}
\left[(\Omega{R})^{(n)}(s)\right]_{\!kj}
\; = \; \sum_{m=1}^{2}\Omega^{nm}_{kk}(s){R}^{(m)}(s)_{kj} \; .
\end{equation}

The final outcome of the calculation then consists in finding the
zeros in the determinant of the $\One\!-(\Omega{R})^{(n)}(s)$ matrix
in Eq.~(\ref{tmatrix}), which give the physical poles in the
complex-$s$ plane.

In a numerical analysis of the poles in the complex plane, we have just considered$k=1$, i.e., when the two
Higgs-field mass states couple only through the $t \bar t$ system.  From very preliminary results, 
the resulting $h$-$H$ mixing is not negligible and tends to reduce the magnitude of
$\Gamma(H\to t \bar t)$. The simpler version of the problem presented in the Appendix contains the main idea. The $h-H$
mixing through $t \bar t$ loops
modifies both real and imaginary parts of the H self-energy.
While the real parts get reabsorbed into the
physical masses, the difference in the imaginary part tends to
reduce the value of $\Gamma(H\to t\bar t)\sim $ 27 GeV, valid for a
conventional 700 GeV Higgs boson, as if the $H t \bar t$ coupling was
smaller than its SM value. This imaginary part, however, vanishes
identically for $s= m^2_h$ and thus this issue is irrelevant
for the on-shell normalization of the 125 GeV resonance
($h-H$ mixing through lighter fermion loops are completely negligible). 

\section{Summary and concluding remarks}

In this paper we started by recalling that, according to
perturbative calculations, the effective potential of the Standard
Model should have a second minimum, well beyond the Planck scale,
which is much deeper than the EW vacuum. Since it is uncertain
whether gravitational effects can become so strong to stabilise the
potential, most authors have accepted the metastability scenario in
a cosmological perspective. This perspective is needed to explain
why the theory remains trapped in our EW vacuum, but requires to
control the properties of matter in the extreme conditions of the
early universe. As a possible alternative, we have then considered
in Sec.~2 the completely different idea of a non-perturbative
effective potential which, as at the beginning of the Standard
Model, is restricted to the pure $\Phi^4$ sector. Nevertheless, the
approximations we have considered have the advantage of being
consistent with two requirements from analytical and
numerical studies of these theories, namely a view of SSB as a weak
first-order phase transition and their basic ``triviality'' in 4D.
In this case, besides the zero-momentum mass $m_h$ defined by the
potential's quadratic shape at the minimum, there is a second,
larger mass scale $M_H$, associated with the ZPE determining the
potential depth. An RG analysis of the effective potential indicates
the existence of two cutoff-independent quantities, namely the mass
$M_H$ itself and a particular definition of the vacuum field, which can be
used to define the Fermi scale $v \sim 246$~GeV. As such, they can
be related by a simple proportionality constant, say $M_H = K v$.
Instead, the smaller mass $m_h$ is cutoff-dependent, because $m^2_h
\sim  M^2_H  L^{-1}$ , in terms of the logarithm of the cutoff $L=
\ln (\Lambda/M_H)$, thus implying the traditional $\Phi^4$ relation $m^2_h
\sim v^2 L^{-1}$. This two-mass structure, obtained from
the effective potential, is consistent with explicit calculations of
the propagator from the Gaussian Effective action (GEA), both for
the one-component and $O(N)$ symmetric theory. In particular the GEA, leading to
the general structure in Eq.(\ref{GEA_general}), explains the large
infrared screening effect for which the quadratic shape $m^2_h$ of
the potential, a genuine zero-momentum quantity, is
much smaller than the large-momentum mass $M^2_H$ governing the size of the
zero-point energy.

The same two-mass structure is found in lattice simulations of the
scalar propagator in the one-component theory, performed to obtain
the best approximations to a free-propagator, in the $p\to 0$ limit
to extract $m_h$, and at larger $p^2$ to extract $M_H$. The
simulations confirm the expected logarithmic scaling trend and allow
to extract the numerical coefficient $c_2$ controlling the
logarithmic slope $M^2_H\sim m^2_h L c_2^{-1}$. In this way, by
combining analytical and numerical indications, one can estimate a
proportionality constant $K \sim 2.80 \pm 0.12$
corresponding to $(M_H)^{\rm Theor} \sim 690(30) $~GeV. This
estimate can also be used to place an upper bound on the lower mass.
Indeed, since $M_H$ is cutoff-independent, when $\Lambda$ gets
smaller the lower mass increases by approaching its upper limit
$(m_h)^{\rm max}\sim M_H$, if $\Lambda$ is as small as possible, say
a few times $M_H$. The resulting limit $(m_h)^{\rm max}\sim
690(30)$~GeV, being in excellent agreement with the average value $(m_h)^{\rm
max} \sim 690\,(50)$~GeV of the traditional theoretical upper bounds
in the $O(4)$ theory, represents a confirmation of our estimate of $M_H$ for the physical
Higgs field. Of course, in the real world $m_h=125$~GeV,
so that, if there was a second resonance with $M_H\sim 690$~GeV, the
ultraviolet cutoff $\Lambda$ should be extremely large.

In Sec.~3, we then argued that, as compared to the effect of such a
large $M_H$, the ZPEs of gauge and fermion fields would represent a
small radiative correction, so that SSB could originate from the
pure scalar sector. A check of our picture is thus demanding the
observation of the second resonance and also its phenomenology. In
this respect, we have observed that, in spite of its large mass, the
second resonance should couple to longitudinal $W$s with the same
typical strength as the low-mass state at 125 GeV and thus represent
a relatively narrow resonance, mainly produced at LHC by gluon-gluon
fusion. For this reason, it is remarkable that, from the LHC data
presented in Sec.~4, one can find combined indications of a new
resonance of mass $(M_H)^{\rm comb} \sim$ 685\,(10)~GeV, with a
statistical significance that is far from negligible. Since this
non-negligible statistical evidence could become an important
discovery with the inclusion of new data, we have also outlined in
Sec.~5 further refinements of the theoretical predictions that could
be obtained by implementing unitarity constraints, in the presence
of fermion and gauge fields, with the type of coupled-channel
calculations used in modern meson spectroscopy (see
Ref.~\cite{PPNP117p103845} for a recent review). In particular, this
method could become useful if the $H$ total width turned out to be
smaller than the expected value $\Gamma_H = 31\div 35$~GeV,
obtained in Sec.~3 from the naive sum of the various partial widths.
In fact, a sizeable defect could possibly indicate the
non-perturbative suppression of some partial width due to the
presence of several competing decay channels and mixing between
$h(125)$ and $H$. A simple argument to illustrate the mixing effect
is presented in the Appendix for the main $\Gamma(H \to t \bar t)$
decay width. 

Before concluding, we will again return to the qualitative
difference that exists between the lattice propagator in the
symmetric phase (see Fig.~\ref{0.074}) and the corresponding plot in
the broken-symmetry phase (see Fig.~\ref{0933}). As we have argued, this
difference can be understood in terms of the peculiar two-mass structure of
the scalar propagator expected on the basis of Gaussian-like
approximations to the effective potential and effective action.

This idea of deviations from a single-mass structure brings us in touch with the work of Van der Bij
\cite{jochum2018}, where a propagator exhibiting a two-mass
structure was considered on the basis of other arguments, quite
unrelated to the effective potential. To this end, he started from
two observations. First, renormalisability does not by itself imply
a single-particle peak, but only a spectral density that falls off
sufficiently fast at infinity. Second, the Higgs field fixes the
vacuum state of the theory which determines the masses of all other
particles. In this sense, the Higgs field itself remains different
and it is not unreasonable to expect deviations of its propagator
from the standard one-pole structure. Here, Van der Bij does not
mention the two-branch spectrum of superfluid He-4, but still the
idea of the SSB vacuum as some kind of medium seems implicit in his
remark. He then considers explicitly the possibility that the
physical Higgs boson is actually a mixture of two states with
Euclidean propagator ($-1\leq\eta\leq 1$)
\begin{equation}
\label{mixing}
G(p) \; \sim \; \frac{1 - \eta}{2} \frac{1}{p^2 + m^2_h } +
\frac{1 + \eta}{2} \frac{1}{p^2 + M^2_H }
\end{equation}
This type of propagator differs from our form in
Eq.~(\ref{GEA_general}) but is a convenient tool to address the issue
of radiative corrections. For instance, let us use Eq.~(\ref{mixing})
in the analysis of the radiative corrections to the $\rho$ parameter
\cite{rho2}, which, in most observables, is the basic quantity to
estimate the virtual effects of the Higgs field. Since the two-loop
correction \cite{jochumtini} is completely negligible for masses
below 1~TeV, one can restrict oneself to the one-loop level, where
the two branches in Eq.~(\ref{mixing}) do not mix, as if one were
replacing in the main logarithmic term an effective mass $m_{\rm
eff} \sim \sqrt{m_h M_H}~(M_H/m_h)^{\eta/2}$. In this way, we could
explore how well the mass parameter $m_{\rm eff}$, as obtained
indirectly from radiative corrections, agrees with $m_h= 125$~GeV,
as measured directly at LHC. Here we just report the two extreme
indications reported in the PDG review \cite{PDG2022}. Namely, from
the experimental set ($A_{\rm LR}$, $M_Z$, $\Gamma_Z$, $m_t$), one
would predict the pair $[m_{\rm eff}=38^{+30}_{-21}$~GeV,
$\alpha_s(M_z)=0.1182\,(47)$]. Alternatively, from the set ($A_{\rm
FB}(b,c)$, $M_Z$, $\Gamma_Z$, $m_t$), one would get the pair
$[m_{\rm eff}=348^{+187}_{-124}$~GeV, $\alpha_s(M_z)=0.1278\,(50)$].

These two extreme cases show that, at this level of precision, one
should estimate the uncertainty induced by strong interactions. This
enters via the contribution of hadronic vacuum polarisation to
$\Delta \alpha(M_z)$, but also more directly through the value of
$\alpha_s(M_z)$. More precisely, in the two examples considered
above, the latter uncertainty enters mainly through $r(M_z)$, i.e.,
the strong-interaction correction to the quark-parton model in
$R(e^+e^- \! \to \! {\rm hadrons})$ at the CM energy $Q=M_z$ (the
quantity that in perturbative QCD is approximated as
$r^{\rm TH}(M_z)=1+\alpha_s(M_z)/\pi + \ldots$). Since, for a given value of
$m_t$, the two quantities  $r(M_z)$ and $m_{\rm eff}$ are positively correlated
in the hadronic $W$ and $Z$ widths, in a global fit this uncertainty will
propagate through these widths and the LEP1 peak cross sections,
thus affecting all quantities, even the pure leptonic widths and
asymmetries.

These positive correlations should not be underestimated because, as pointed
out in Ref.~\cite{fiore1992}, there is a sizable excess in the
$(e^+e^- \to {\rm hadrons})$ data. As a consequence, to obtain from the PETRA,
PEP, and TRISTAN data the correct position of the $Z$ peak, one has to
substitute in $r^{\rm TH}$ (34~GeV) a value $\alpha_s (34 {\rm GeV}) \sim 0.17$
that is considerably larger than the canonical 0.14 predicted from deep
inelastic scattering. At a later date, some excess was also observed at LEP2
\cite{wynhoff,wilkinson,erler}.
The whole issue of $R(e^+e^- \! \to \! {\rm hadrons})$ was
reconsidered by Janot \cite{janot} in a thorough analysis of all data from
20 GeV up to LEP1 and from LEP1 up to LEP2. In this analysis, at each CM
energy, one introduces a small correction $\delta$ through the relation
\BE
R^{\rm EXP} \;  = \; R^{\rm SM} (1 + \delta) \; ,
\EE
determining average values from the data below LEP1 and from the data above
LEP1, namely $\delta_{\rm low}=0.0079\,(52)$ and $\delta_{\rm high}=0.015\,(9)$,
respectively. Therefore, by extrapolating, from both sides, to the $Z$ peak,
we find a combined value $\delta_{\rm comb}=$ 0.0097\,(45). To understand the
implications of this tiny one-percent correction, let us compare with the SM
predictions. In Janot's analysis, $R^{\rm SM}$ was computed with
a QCD correction producing $\alpha_s (M_z)=0.1183\,(20)$. This gives a leading
non-singlet correction, being the same for $\gamma$ and $Z$ exchange,
$r^{\rm TH}(M_z) \sim 1+\alpha_s(M_z)/\pi=1.03766\,(64)$, in very good
agreement with the central value $\langle r^{\rm TH}(M_z) \rangle=1.03756$
obtained, with $ \alpha_s (M_z) =$ 0.1181, in the recent four-loop computation
of Ref.~\cite{cinesi}. Now, with the PDG $Z$ hadronic width
$\Gamma(Z \to {\rm hadrons})=1744.4 \pm 2.0$~MeV, we can check the effect of
$\delta_{\rm comb}$ by computing the bare width in the quark-parton model
$\Gamma(Z \to q \bar q)$ as
\begin{eqnarray}
\Gamma(Z \to q \bar q) & = &
\frac{\Gamma(Z \to {\rm hadrons})}{(1+\delta_{\rm comb})(1+0.03766\,(64))}
\; \sim \; \nonumber \\[1mm]
&& (1665 \pm 8 ) \; {\rm MeV} \; .
\end{eqnarray}
The resulting value would thus be about 2$\sigma$ smaller than the estimate
$\Gamma(Z \to q \bar q)= 1681.3$ MeV reported in Ref.~\cite{cinesi} for
$m_t=172.9$~GeV and $m_h=125$~GeV. For the same top quark mass, a bare width
1665\,(8)~MeV would require an effective Higgs mass in radiative corrections
that is considerably larger than the mass 125~GeV measured at LHC. But then
this would produce a tension with the PDG leptonic width
$\Gamma(Z \to l^+l^-)=83.984\,(86)$~MeV, because the theoretical prediction
for the ratio $R^{(0)}=\Gamma(Z \to q \bar q) /\Gamma(Z \to l^+l^-)$ is nearly
insensitive to the precise values of $m_t$ and $m_h$. Our point is that the
small excess in $R(e^+e^- \to {\rm hadrons})$, observed at 34~GeV and which
persists up to LEP2, intertwines with and influences a precision test of the
pure electroweak corrections. For this reason, the present view that the Higgs
mass parameter extracted indirectly from radiative corrections agrees perfectly
with the $m_h=125$~GeV measured directly at LHC is not free of ambiguities.

\begin{acknowledgement}
We thank Leonardo Cosmai and Fabrizio Fabbri for many useful discussions
and collaboration. We also thank Patrick Janot for information on his analysis
of the $ e^+e^- \to\  {\rm hadrons}$ data.
\end{acknowledgement}

\appendix
\section{\boldmath{$\Gamma(H \! \to \! t \bar t)$} reduction due to
\boldmath{$h\!-\!H$} mixing}
Here we will give a simple argument to expect that, as an effect of
$h$-$H$ mixing, the  $\Gamma(H \to t \bar t)$ decay width will be smaller
than the traditional estimate for a Higgs resonance with the same mass.
The starting point is the basic one-loop function with a $t \bar t$ pair:
\begin{equation}
\Omega(s) \; = \; \frac{N_c m^2_t}{v^2}\frac{1}{(2\pi)^4}
\left[4 A(m_t)+(2s-8m^2_t) B_0(s)\right] \; ,
\end{equation}
where
\begin{equation}
A(m_t) \; = \;  \int \! \frac{d^4q}{q^2 + m^2_t}
\end{equation}
and
\begin{equation}
B_0(s) \; = \; \int \! d^4q \frac{1}{q^2 + m^2_t} \frac{1}{(q+P)^2 + m^2_t} \;,
\end{equation}
with $P^2=-s$. By using dimensional regularisation, in the
$\overline{MS}$ scheme with 't Hooft scale $\mu$, we then find
\begin{equation}
\Omega(s) \; = \; \frac{N_c}{8\pi^2}\frac{m^2_t}{v^2}
\left[2a(m_t) + (s-4m^2_t) \, b_0(s)\right] \; ,
\end{equation}
with
\begin{equation}
a(m_t) \; = \; -2 m^2_t \left[\ln \frac{\mu}{m_t} + \frac{1}{2}\right]
\end{equation}
and
\begin{equation}
b_0(s) \; = \; 2\ln \frac{\mu}{m_t}  + 2 -
\beta_t \ln  \frac{1+\beta_t}{1-\beta_t}  + i\pi\beta_t \; ,
\end{equation}
where $\beta_t= \sqrt {1-4m^2_t/s}$ and we have assumed $s > 4m^2_t$.
By expressing
\begin{equation}
\Omega(s) \; = \; \Omega_R(s) + i \Omega_I(s) \; ,
\end{equation}
we then find
\begin{equation}
\Omega_I(s) \; =  \; \frac{N_c}{8\pi} \frac{m^2_t}{v^2} \, s \, \beta^3_t
\end{equation}
and
\begin{equation}
\label{first}
\Omega_R(s) \; = \; \frac{N_c}{8\pi^2} \frac{m^2_t}{v^2} \, s \,
\left\{ 2\left[1-\frac{6m^2_t}{s}\right] \ln \frac{\mu}{m_t} +f_0(s)\right\}\;,
\end{equation}
where
\begin{equation}
f_0(s) \; = \; 2 \left[1-\frac{5m^2_t}{s}\right] - \beta^3_t \,
\ln \frac{1+\beta_t}{1-\beta_t} \; .
\end{equation}
By including $h$-$H$ mixing, we then get the series
\begin{eqnarray}
\hat \Omega(s) & = & \Omega(s) \, + \, \Omega(s) \, G_h(s) \, \Omega(s) \, +
\nonumber \\[0.5mm]
&&\Omega(s) \, G_h(s) \, \Omega(s) \, G_h(s) \, \Omega(s) \, + \, \ldots \, =
\nonumber \\
&&\frac{\Omega(s)} {1-\Omega(s) G_h(s)}  \; ,
\end{eqnarray}
where the light-Higgs propagator is $G_h(s)=1/(-s + m^2_h)$ and purely real.
By introducing $\bar s= M^2_H -i M_H\Gamma_H$ to indicate the location of the
pole, while defining $\bar \Omega_R \equiv \mbox{$\cal R$e}(\Omega(\bar s))$
and $\bar\Omega_I\equiv\mbox{$\cal I$m}(\Omega(\bar s))$,  we obtain, for
$m^2_h \ll M^2_H$,
\begin{equation}
M^2_H - i M_H\Gamma_H \; = \;  M^2_B - \frac{\bar\Omega_R +i \bar \Omega_I}
{ 1 + \frac{\bar\Omega_R +i \bar \Omega_I}{M^2_H} } \; ,
\end{equation}
or
\begin{equation}
M^2_H \; = \;  M^2_B - \frac {\bar\Omega_R ( 1 + \frac{\bar\Omega_R} {M^2_H}) +
\frac{\bar \Omega^2_I}{M^2_H}}  { (1 + \frac{\bar\Omega_R} {M^2_H})^2 +
\frac {\bar \Omega^2_I}{M^4_H}  }
\end{equation}
and
\begin{equation}
M_H\Gamma_H \; =  \; \frac{\bar\Omega_I}{(1 +\frac{\bar\Omega_R} {M^2_H})^2 +
\frac {\bar \Omega^2_I}{M^4_H}  }  \; .
\end{equation}

To eliminate formally the dependence on the scale $\mu$, we can then introduce
the mass parameter $\Delta= M_B -M_H$, with $M_B$ the bare heavy-Higgs mass,
in terms of which, by neglecting $\bar \Omega^2_I/M^4_H$, we find
\begin{equation}
\bar\Omega_R \; \sim \; \frac {2M_H \Delta } { 1 -  \frac{2\Delta}{M_H}  }
\end{equation}
and
\begin{equation}
\Gamma_H\;\sim\;\frac{\bar\Omega_I}{M_H}\,\left(1-\frac{4\Delta}{M_H}\right)\;.
\end{equation}
Therefore, as an effect of the $h$-$H$ mixing, the physical width
$\Gamma_H=\Gamma(H \to t \bar t)$ is reduced with respect to its traditional
value $\bar\Omega_I/M_H $, provided that $\Delta/M_H > 0$. In turn, a value
$\Delta/M_H > 0$ implies $\bar\Omega_R/M^2_H>0$ or, from Eq.~(\ref{first}),
\begin{equation}
\left\{2\left[1-\frac{6m^2_t}{M^2_H}\right] \ln \frac{\mu}{m_t} +
f_0(M^2_H)\right\} \; > \; 0 \; .
\end{equation}
Now, for $\mu\sim M_H \sim 690 $ GeV, the value of the above quantity is
1.246 $\ln (M_H/m_t) -0.33 = 1.39$. Thus, $\Delta/M_H >0$ is the natural
choice. Moreover, larger values of $\mu$ cannot be excluded, because the
loop is actually an ultraviolet divergent expression. In the usual one-loop
calculation, the problem does not arise because the imaginary part is finite
and the divergence is reabsorbed into an unobservable renormalisation of the
bare mass. But here, through the infinite chain of the $h$-$H$ mixing diagrams,
the ultraviolet divergence in the real part affects the imaginary part, which
may exhibit a potentially large defect. However, this defect will be mitigated
in higher orders, where the constant $m_t$ in the main vertex of the one-loop
calculation is replaced by a running mass $m_t(\mu)$, which follows the
logarithmic decrease of the corresponding Yukawa coupling.

Of course, for consistency, the physical mass $M_H$ should also be smaller
than the bare-mass parameter $M_B$. This is more difficult to check, for two
reasons. First, $\Delta$ is here just a theoretical parameter that only takes
into account the real part of the $t \bar t$ loop. To consider the experimental
quantity $\Delta^{\rm EXP} = M_B - M^{\rm EXP}_H$, one should include all
self-energy parts. Second, even by restricting ourselves to the $t \bar t$
loops, the theoretical uncertainty of our prediction $M_B\sim 690\pm 30$~GeV
is non-negligible. In practice, any $M_H$ that is smaller than 720~GeV should
be allowed.

In conclusion, together with the definite mass range and the sharp
$\gamma_H$-$\sigma_R$ correlation in the charged four-lepton channel, this
effect of $h$-$H$ mixing on the $H \to t \bar t$ width represents a third
characteristic prediction of our picture.

\end{document}